# Phenomenology with Wilson fermions using smeared sources

David Daniel and Rajan Gupta

*T-8, MS-B285, Los Alamos National Laboratory, Los Alamos, NM 87545*

Gregory W. Kilcup

*Physics Department, The Ohio State University, Columbus, OH 43210*

Apoorva Patel

*Supercomputer Education and Research Centre and Centre for Theoretical Studies*
*Indian Institute of Science, Bangalore 560012, India*

Stephen R. Sharpe*

*Theory Group, CEBAF, Newport News, VA 23606*

We investigate the use of two types of non-local ("smeared") sources for quark propagators in quenched lattice QCD at $\beta = 6.0$ using Wilson fermions at $\kappa = 0.154$ and 0.155. We present results for the hadron mass spectrum, meson decay constants, quark masses, the chiral condensate and the quark distribution amplitude of the pion. The use of smeared sources leads to a considerable improvement over previous results. We find a disturbing discrepancy between the baryon spectra obtained using Wuppertal and wall sources. We find good signals in the ratio of correlators used to calculate the quark mass and the chiral condensate and show that the extrapolation to the chiral limit is smooth.

---

*On leave from Physics Department, University of Washington, Seattle, WA 98195

## 1. Introduction

In addition to reducing systematic errors in lattice QCD due to finite lattice volume, finite lattice spacing and extrapolation from heavy quark masses to physical $m_u$ and $m_d$, it is important to improve the signal in observables. The particular observables we are interested in are matrix elements within hadronic states; the corresponding correlation functions from which these are extracted are made up of products of quark propagators. Even though at present the quark propagators are calculated without specific information on the motion of the other valence quarks in a given bound state, it has been shown that the signal is significantly improved by using smeared quark sources [1] [2].

A number of different smearing methods have been investigated and some of these have been reviewed in Ref. [3]. Our choices here are motivated by the requirements of a parallel calculation of weak matrix elements using Wilson fermions [4]. Our calculation requires two kinds of hadron sources: one that produces hadrons with zero momenta and the other that couples to all momenta. We construct zero momentum hadron correlators using "wall" source propagators while the "Wuppertal" source propagators [2] yield hadron correlators that have overlap with all momenta. In this paper we demonstrate the efficacy of these two kinds of correlators by investigating the signal in both the amplitudes and the masses obtained from 2-point correlation functions. The improvement is exemplified by the results obtained for the spectrum and meson decay constants.

We do, however, find a disconcerting difference between the masses of baryons extracted using Wuppertal and wall sources. Even though at this stage the difference between the central values is 1–3 $\sigma$, it has a significant effect on the nucleon to rho mass ratio.

An important prerequisite when using Wilson fermions to calculate matrix elements is an understanding of the realization of chiral symmetry. It was shown in Refs. [5] and [6] that the conventional continuum current algebra can be reproduced on the lattice provided suitably modified chiral Ward identities are used. Among other things, this requires multiplying lattice fermion bilinears by renormalization constants to relate them to their continuum counterparts. We use the following definitions

$$\begin{aligned}
S^{\rm cont}(x) &= Z_S S(x); & S(x) &= \overline{\psi}(x)\psi(x), \\
P^{\rm cont}(x) &= Z_P P(x); & P(x) &= \overline{\psi}(x)\gamma_5\psi(x), \\
V_\mu^{\rm cont}(x) &= Z_V V_\mu(x); & V_\mu(x) &= \overline{\psi}(x)\gamma_\mu\psi(x), \\
A_\mu^{\rm cont}(x) &= Z_A A_\mu(x); & A_\mu(x) &= \overline{\psi}(x)\gamma_\mu\gamma_5\psi(x).
\end{aligned} \tag{1.1}$$

If we were working with a chirally invariant regulator and a mass independent continuum renormalization scheme, we would have $Z_V = Z_A = 1$ and $Z_S = Z_P = 1/Z_{\rm mass}$, where $Z_{\rm mass}$ is the mass renormalization. The latter three constants have anomalous dimensions and therefore depend on the momentum scale. On the lattice with Wilson fermions $Z_V$ and $Z_A$ are finite renormalization constants and $Z_P \neq Z_S$. Although all these constants can be calculated in perturbation theory (see Appendix A), it appears that for present couplings ($g^2 \approx 1$) there are large non-perturbative contributions. For $Z_A$ and $Z_V$, non-perturbative calculations are possible, and have been performed by a number of authors. These estimates include, in general, large $O(a)$ effects, and therefore different non-perturbative techniques can give different estimates. We use the estimates most appropriate for our study, $Z_A = 0.86$ and $Z_V = 0.57$ [7].

The constants $Z_S$ and $Z_P$ are more problematic because, unlike $Z_A$ and $Z_V$, they depend upon the choice of continuum renormalization scheme. We need to determine these constants if we are to relate the cutoff dependent lattice results for the chiral quantities $m_s$ and $\langle\overline{\psi}\psi\rangle_{m_q=0}$ to the scale dependent continuum values. A convenient continuum scheme is the $\overline{\rm MS}$ scheme at $\mu = 1$ GeV. Since this is a perturbative scheme, the most natural thing to do is to use the lattice perturbative results for $Z_S$ and $Z_P$. These are discussed in Appendix A. If perturbation theory is accurate (which is unlikely in the present simulation) then one can make unambiguous predictions for $m_s$ and $\langle\overline{\psi}\psi\rangle_{m_q=0}$ in the continuum.

Alternatively, one can proceed in a phenomenological fashion and demand, for example, that the lattice results reproduce the value of $m_s$ in the $\overline{\rm MS}$ scheme. With this assumption $Z_P$ is fixed, and the Ward identities then determine $Z_S$ non-perturbatively. The question is then one of consistency: with $Z_S$ and $Z_P$ fixed in this way, can all lattice quantities whose definitions depend upon $Z_S$ and $Z_P$ be made to agree with the $\overline{\rm MS}$ scheme? In quenched QCD we can expect this to be only approximately true. We address these questions further in Secs. 7 and 9. For the present we simply state that our data do favor values of $Z_P$ and $Z_S$ very different from their perturbative values.

This paper is organized as follows: in Sec. 2 we describe the lattices used and our method for extracting physical parameters from 2-point correlators. The types of quark sources used to construct the hadron correlators are defined in Sec. 3. We present an analysis of the spectrum in Sec. 4 and of the meson decay constants, $f_\pi$ and $f_V^{-1}$, in Secs. 5 and 6. The calculation of the current quark mass and chiral condensate is presented in Sec. 7, and of the second moment of the pion distribution amplitude in Sec. 8. The extrapolation to the chiral limit is discussed in Sec. 9, and we end with conclusions in Sec. 10. The perturbative results for the renormalization constants are given in Appendix A.

## 2. Lattice parameters and fitting procedure

Our statistical sample consists of 35 lattices of size $16^3 \times 40$ at $\beta = 6.0$, generated as two independent streams. The first stream consists of 14 lattices, generated using a pseudo heatbath algorithm, which are separated by 1000 sweeps. The second stream consists of 21 lattices generated using a combination of over-relaxed and Metropolis algorithms, and are separated by 300 combination sweeps. These sets of lattices have been used previously for spectrum and weak matrix element analysis using both Wilson and staggered fermions [8].

The major thrust of this study is to explore improved numerical techniques for Wilson fermions. For this purpose we use only two values of the quark mass, $\kappa = 0.154$ and $0.155$. Translated to physical units these correspond to pion mass values of roughly 660 MeV and 540 MeV, respectively. The criterion used to judge whether the quark propagator has converged is $R = |\frac{M\chi - \phi}{\chi}|^2$, where $\chi$ is the solution, $\phi$ is the source vector and $M$ is the Wilson operator. We find no significant difference in the long time tail of propagators when reducing $R$ from $10^{-15}$ to $10^{-18}$, at either quark mass. To be conservative, we adopt $R = 3 \times 10^{-16}$ as our convergence criterion.

To extract the amplitudes and masses from the long time behavior of the 2-point correlators, we make fits assuming that the lowest mass state dominates the correlation function. To ensure this we first examine the effective mass plot for a plateau and then make a single mass fit selecting the range of the fit based on the following criteria: (a) $t_{\rm min}$ always lies in the plateau, (b) $t_{\rm max}$ is selected to be as large as possible consistent with a signal. In most cases we find that the central value obtained from the fits is the same with and without using the full covariance matrix. In some cases we cannot use the full range of the plateau because the covariance matrix is close to being singular. In such cases the problem is not that we cannot invert the covariance matrix but that the result is very sensitive to the range of the fit (the central value can change by one or more standard deviations on the addition of a single point to the fit range). Our tests using subsets of the data show that this instability is a result of inadequate statistics. We, therefore, reduce the range of the fit to get a stable result. For example, even though the signal in the pion correlators lasts until $t \approx 40$, we use $t_{max} \le 25$ in order to avoid fits based on almost singular covariance matrix.

All error estimates are obtained using a single-elimination jack-knife procedure. Previous analysis [8] [9] leads us to believe that the lattices used are sufficiently decorrelated for this method to be adequate.

## 3. Quark propagators and hadron correlators

The calculation of quark propagators is done on lattices doubled in the time direction, *i.e.* $16^3 \times 40 \rightarrow 16^3 \times 80$. We use periodic boundary conditions in all four directions. These propagators on doubled lattices are identical to a linear combination of propagators calculated with periodic ($P$) and antiperiodic ($A$) boundary conditions on the original $16^3 \times 40$ lattice. For the source on time slice 1, the forward moving solution (time slices 2–40) is $F = (P + A)/2$ while the backward moving solution (time slices 80–42) is $B = (P - A)/2$. We find no numerical difference between calculating $F$ and $B$ directly or from the combination of $P$ and $A$. We use doubled lattices because our weak matrix element calculations require both forward and backward moving solutions. We will show that doubling also helps in the spectrum analysis.

We calculate quark propagators using two different types of smeared sources: "Wuppertal" and "wall". The wall source for a given spin and color consists of a delta function at each point on time slice $t = 1$, with the wall time slice fixed to Coulomb gauge. The hadron correlators built from these propagators have overlap with only zero-momentum states.

The Wuppertal source is the covariant solution of the 3-dimensional Klein-Gordon operator $K$ with a delta function source at $x = y = z = 1$ on time slice $t = 1$

$$K(x;y) = \delta_{x,y} - \kappa_{KG} \sum_{i=1}^{3} \left( U(x,i)\delta_{x+i,y} + U^\dagger(x,i)\delta_{x-i,y} \right) , \qquad (3.1)$$

where the parameter $\kappa_{KG}$ controls the size of the smearing. The implementation of this smearing method is the same as that used in Ref. [10], and the details have been presented there. We choose $\kappa_{KG} = 0.181$, for which the size of the smeared source (defined in Ref. [10]) varies between 4.1–4.5. A statistical problem when using the Wuppertal source is that, as one increases $\kappa_{KG}$, the configuration to configuration fluctuations in the hadron correlators increase. For our value of $\kappa_{KG}$, the fluctuations at long time separation are less than a factor of 5.

We also construct a third kind of quark propagator which is smeared at both the source and sink points. This is obtained by applying the inverse of the Klein-Gordon operator to the result of Wuppertal source propagator at each time slice.

From the three kinds of quark propagators we construct three kinds of hadron correlators: (a) wall source and local (point) sink (labeled henceforth as $LW$), (b) Wuppertal smeared source and local sink ($LS$) and (c) Wuppertal smeared source and sink ($SS$). We use the following notation to define the 2-point hadron correlators:

$$\begin{aligned}
\Gamma_{LW}(t) &= \langle 0|\mathcal{O}^{\rm local}(t) J^{\rm wall}(0)|0\rangle \overset{t\to\infty}{\sim} \frac{\langle 0|\mathcal{O}^{\rm local}|h\rangle\langle h|J^{\rm wall}|0\rangle}{2M} e^{-Mt} , \\
\Gamma_{LS}(t) &= \langle 0|\mathcal{O}^{\rm local}(t) J^{\rm smeared}(0)|0\rangle \overset{t\to\infty}{\sim} \frac{\langle 0|\mathcal{O}^{\rm local}|h\rangle\langle h|J^{\rm smeared}|0\rangle}{2M} e^{-Mt} , \\
\Gamma_{SS}(t) &= \langle 0|\mathcal{O}^{\rm smeared}(t) J^{\rm smeared}(0)|0\rangle \overset{t\to\infty}{\sim} \frac{\langle 0|\mathcal{O}^{\rm smeared}|h\rangle\langle h|J^{\rm smeared}|0\rangle}{2M} e^{-Mt},
\end{aligned} \qquad (3.2)$$

where $|h\rangle$ is the appropriate hadronic state that saturates the 2-point correlator at large $t$ and $M$ its mass. The operator $J$ creates the hadron and is always constructed from smeared (Wuppertal or wall) quark sources in our calculation, and $\mathcal{O}$ is the operator used to destroy it. The projection onto a definite momentum state is always done at the sink time slice. We compare the efficacy of these correlators for the three types of quark sources in the next section.

## 4. Spectrum

The flavor non-singlet meson and the baryon interpolating operators we have used are as follows:

$$\begin{aligned}
\pi &= \overline{\psi}\gamma_5\psi \\
\pi_2 &= \overline{\psi}\gamma_4\gamma_5\psi \;\equiv\; A_4 \\
\rho &= \overline{\psi}\gamma_i\psi \\
\rho_2 &= \overline{\psi}\gamma_4\gamma_i\psi \\
a_0 &= \overline{\psi}\psi \\
a_1 &= \overline{\psi}\gamma_i\gamma_5\psi \\
b_1 &= \overline{\psi}\sigma_{ij}\psi \\
N^{1/2+} &= \epsilon_{abc}\big(u^a\mathcal{C}\gamma_5 d^b\big)u_1^c \quad \text{and} \\
&= \epsilon_{abc}\big(u^a\mathcal{C}\gamma_5 d^b\big)u_2^c \\
\Delta_1^{3/2+} &= \epsilon_{abc}\big(u^a\mathcal{C}\frac{\gamma_1 - i\gamma_2}{2}u^b\big)u_1^c \quad \text{and} \\
&= \epsilon_{abc}\big(u^a\mathcal{C}\frac{\gamma_1 + i\gamma_2}{2}u^b\big)u_2^c \\
\Delta_2^{3/2+} &= \epsilon_{abc}\big(u^a\mathcal{C}\gamma_3 u^b\big)u_1^c - \epsilon_{abc}\big(u^a\mathcal{C}\frac{\gamma_1 - i\gamma_2}{2}u^b\big)u_2^c \quad \text{and} \\
&= \epsilon_{abc}\big(u^a\mathcal{C}\gamma_3 u^b\big)u_2^c + \epsilon_{abc}\big(u^a\mathcal{C}\frac{\gamma_1 + i\gamma_2}{2}u^b\big)u_1^c,
\end{aligned} \tag{4.1}$$

where $\mathcal{C} = \gamma_4\gamma_2$ is the charge conjugation matrix. We specify the total angular momentum and the parity for the baryons (the opposite parity baryons are obtained from the backward moving solutions using the same operators). For the $\Delta$ we use two different Lorentz structures ($\Delta_1^{3/2+}$ has $J_Z = \pm 3/2$ while $\Delta_2^{3/2+}$ has $J_Z = \pm 1/2$) to check for systematic errors. The correlators for the two $J_Z$ terms are averaged to improve the statistics. The subscript on $u^c$ specifies the Dirac index of the spinor. The results for the masses in lattice units along with the range and $\chi^2/N_{\rm DF}$ of the fit are given in Tables 1-4. The effective mass plots for each of the three sources at $\kappa = 0.155$ are shown in Figs. 1a, b and c. The corresponding plots at the heavier quark mass, $\kappa = 0.154$, have smaller errors.

We find that the effective mass, $m_{\rm eff}(t)$, converges from above for $SS$ (as it must with a positive definite transfer matrix) and for $LS$ correlators, and from below for the $LW$ correlator. An agreement between the different methods therefore provides a good test of whether the mass estimate is asymptotic.

In the pion channel we find that $m_{\rm eff}(t)$ reaches a plateau at $t = 10$ for $LS$ correlators, at $t = 8$ for $SS$ correlators, while the $LW$ correlators stabilize even earlier at $t = 6$. The onset of the plateau is independent of the quark mass in the two cases considered. In the plateau region the three different sets of correlators give consistent results. The signal in all three cases extends all the way across the lattice and the errors stay roughly constant with $t$. A theoretical analysis of why this occurs for the pion and not for other states has been given by Lepage [11]. The errors in $m_{\rm eff}(t)$ are comparable for $LS$ and $LW$ correlators and roughly a factor of 2 larger in $SS$. In Ref. [10] we gave an argument for why, with Wuppertal smearing, the errors in $\Gamma_{SS}$ are larger than those in $\Gamma_{LS}$. On the other hand, an analysis of the eigenvalues of the covariance matrix shows that the time slice to time slice correlations in $\Gamma_{SS}$ are significantly smaller. The interplay of these two effects makes the errors in the fit parameters comparable in the two cases.

Within the plateau region we have investigated the dependence of the fit parameters on the fit range used and on whether or not we use the full covariance matrix. We find that while the variation in $M_\pi$ is no greater than 1%, the result for the amplitude can change by as much as 12%. This is one reason why the extraction of matrix elements is prone to large errors.

We find that the operator $\pi_2$ has a better overlap with the pion and $m_{\rm eff}(t)$ reaches a plateau earlier. On the other hand the errors in $m_{\rm eff}(t)$ are much larger. A comparison of signals for the two channels can be made from Figs. 1b and 2a.

In Fig. 2a we plot $m_{\rm eff}(t)$ for the four lowest momentum states coupled to the $A_4A_4$ $LS$ correlator. It is desirable to get a good signal in non-zero momentum channels as these correlators play an important part in the calculation of many matrix elements, for example structure functions and form factors. The data show a reasonable plateau for $\vec{p} = (0, 0, 2\pi/16)$ and the possible beginning of one for $\vec{p} = (0, 2\pi/16, 2\pi/16)$. We compare the resulting energy estimates with the continuum dispersion relation in Fig. 2b.

The signal in the rho correlators is also very satisfactory. There is a plateau between $t = 10$ and $t = 30$ for all three types of correlators, but unlike the case of the pion the errors do increase with $t$. The final estimates from the three sets are consistent. We find that the signal is a little better for the $\rho$ than for the $\rho_2$ operator.

There is a reasonable signal in the nucleon ($N^{1/2+}$) channel with a plateau extending between $t = 10$ and 18. We find that the plateau in the $\Delta_2^{3/2+}$ channel has a marginally better signal than in the $\Delta_1^{3/2+}$, and for this reason we quote results using $\Delta_2$. The big surprise is the systematic difference in the baryon spectrum as extracted using the Wuppertal and wall sources. In Fig. 3 we show an effective mass plot comparing the $LS$ and $LW$ correlators for both nucleon and $\Delta_2$. (We emphasize that either Wuppertal correlator, $SS$ or $LS$, could equally well have been used for the comparison with $LW$.) Both types of correlators appear to reach plateaus, though from opposite directions. The problem is that the asymptotic values differ, the $LW$ results lying below the $LS$. The same is true for both values of the quark mass. To explore the statistical significance of this difference we performed a jacknife analysis of the difference of the masses. We find that the difference for the nucleon is only 1 $\sigma$, while that for the $\Delta$ is 2–3 $\sigma$. Even though the effect is of marginal significance in each channel, the fact that in all three baryon channels the $LW$ masses lie below the $LS$ and $SS$ masses suggests that the difference is systematic.

Further evidence for this conclusion comes from a comparison of our results with those from the APE collaboration [12], both of which are collected in Table 5. (Our results use the operators $\pi$, $\rho$, $N$ and $\Delta_2$.) The APE collaboration uses yet another type of source, the "multicube", which is, in a sense, intermediate between wall and Wuppertal sources. The meson masses are in good agreement, while the APE results for baryons lie in between ours for Wuppertal and wall sources.

A similar effect has been observed in hadron mass calculations using staggered fermions [13] [14]. Meson masses agree, while baryon mass estimates with multicube source again lie a couple of standard deviations higher than those with wall sources.

If this is a systematic effect, it means that the extracted mass estimate is not asymptotic. Any such difference has important implications for the $M_N/M_\rho$ ratio. Calculations over the last two years have found a smaller value for this ratio than previous lattice results. This improvement has come, however, from calculations that use wall or similar sources [3]. Clearly this is an issue deserving further study.

For the positive parity mesons and the negative parity baryons the signal is marginal. In most cases the plateau is noisy and extends only over $\sim 5$ time slices. For this reason we give the results of the fits but do not present a detailed analysis of the data.

The signal in the $\pi$ and $\rho$ channels extends beyond half the length of the original gauge configuration ($t = 20$), so we can address the question raised by the HEMCGC collaboration about fluctuations in the effective mass induced by doubling the lattices [15]. The plateaus in our $m_{\rm eff}(t)$ plots do show fluctuations, but these are not correlated with the lattice periodicity. Furthermore, if we analyze different subsets of lattices we do not find fluctuations at the same values of $t$. In addition, we have calculated the spectrum using quark propagators on the undoubled $16^3 \times 40$ lattices with periodic boundary conditions in the time direction. These results are included in Table 5. There is no significant difference between the mass estimates from doubled or undoubled lattices. In sum, our data shows no evidence that the fluctuations are anything other than statistical. It remains possible, however, that systematic fluctuations would appear if the statistical errors were reduced.

## 5. Pseudo-scalar decay constant $f_\pi$

The definition of the pseudo-scalar decay constant on the lattice is

$$f_\pi = \frac{Z_A \langle 0 | A_4^{\rm local} | \pi(\vec{p}) \rangle}{E_\pi(\vec{p})} \ , \tag{5.1}$$

where $Z_A$ is the axial current renormalization, and we are using the convention that the experimental value is $f_\pi = 132$ MeV. In order to extract $f_\pi$ we study two kinds of ratios of correlators:

$$
\begin{aligned}
R_1(t) &= \frac{\Gamma_{LS}(t)}{\Gamma_{SS}(t)} \quad \overset{t\to\infty}{\sim} \quad \frac{\langle 0|{A_4}^{\rm local}|\pi\rangle}{\langle 0|{A_4}^{\rm smeared}|\pi\rangle} \\
R_2(t) &= \frac{\Gamma_{LS}(t)\Gamma_{LS}(t)}{\Gamma_{SS}(t)} \quad \overset{t\to\infty}{\sim} \quad \frac{|\langle 0|A_4^{\rm local}|\pi\rangle|^2}{2M_\pi}\; e^{-M_\pi t},
\end{aligned}
\tag{5.2}
$$

where the $\Gamma$ are defined in Eq. (3.2). In the case of $R_1$ we have to extract $\langle 0|{A_4}^{\rm smeared}|\pi\rangle$ separately from the $\Gamma_{SS}$ correlator. For each of the two ratios, $R_1$ and $R_2$, the smeared source $J$ used to create the pion can be either $\pi$ or $\pi_2$. Thus we calculate $f_\pi$ in four different ways, which we label as $f_\pi^a$ (using ratio $R_1$ with $J = \pi$), $f_\pi^b$ (using ratio $R_1$ with $J = \pi_2$), $f_\pi^c$ (using ratio $R_2$ with $J = \pi$), and $f_\pi^d$ (using ratio $R_2$ with $J = \pi_2$). The results are given in Table 6. Errors are estimated by applying single elimination jackknife to the estimate of $f_\pi$ itself. We note in passing that the "naive" estimate obtained by combining the errors in each of the fit parameters in quadrature results in errors that agree with quoted errors to better than 10%.

We find that all four methods give consistent estimates of $f_\pi$. Since we have to combine different correlators in order to extract $f_\pi$ we select the fit range based on the following criteria: (1) goodness of the fit, (2) presence of a plateau with a similar mass estimate from each of the individual correlators. Otherwise, as stated before, there can be a large variation in the estimate of the amplitude. The quality of the signals for the correlator ratios leading to $f_\pi^a$ and $f_\pi^c$ are exemplified by Figs. 4a and 4b.

Bernard *et al.* [16] have calculated $f_\pi$ using point sources at the same two values of $\kappa$ and using a sub-set of the lattices analyzed by us (they use every other lattice). They get $f_\pi = 0.094(7)$ and $0.086(9)$ at $\kappa = 0.154$ and $0.155$ respectively, in good agreement with our values. We also find agreement with recent point-source results of the QCDPAX collaboration [17]. The results of the APE collaboration [12] are in error by a factor $\sqrt{e^{-M_\pi}}$ [18]. After correcting for this factor their value at $\kappa = 0.155$ (obtained using extended cube sources) is consistent with our corresponding result $f_\pi^d$.

We have also calculated $f_\pi$ at non-zero momentum, and find that its value is stable to the addition of one unit of momentum. The results are included in Table 6. The overall quality of the signal is good even though the plateaus in the effective mass are much shorter than those for the $\vec{p} = (0,0,0)$ case. At $\vec{p} = (0, 2\pi/16, 2\pi/16)$ the errors are larger and the determination is far less reliable.

## 6. Vector decay constant $f_V$

We use the local vector current $V_i$ to define the dimensionless number $f_V^{-1}$ as

$$
\langle 0|V_i|\rho\rangle = \frac{\epsilon_i M_V^2}{Z_V f_V}. \tag{6.1}
$$

This matrix element can be extracted in two ways analogous to Eq. (5.2). We use $V_i$ to both create and annihilate the vector meson. The results are given in Table 7, where we use the non-perturbative estimate $Z_V = 0.57$ obtained using 2-point correlators in an earlier calculation of $f_V$ at the same $\beta$ [7]. We point out that due to the existence of a conserved vector current on the lattice, this estimate of $Z_V$ is free of $O(g^2)$ ambiguities.

The quality of the signal in correlator ratios $R_1$ and $R_2$ is very good as shown in Figs. 5a and 5b. The final data is shown in Fig. 6 where for comparison we have also included results from the Wuppertal [19], APE [12] and QCDPAX collaborations [17]. The experimental points have been taken from Ref. [19]. Our results lie significantly below Wuppertal estimates and are in agreement with the results obtained by the QCDPAX collaboration and the APE collaboration (the latter after correction by a factor $\sqrt{e^{-M_\rho}}$ [18]). The data by the QCDPAX collaboration, obtained using point source propagators measured on 160 lattices of size $24^3 \times 54$, show that smeared and local sources yield consistent results once systematic errors are under control.

## 7. Quark masses and the chiral condensate

In order to extract the quark masses and the condensate with Wilson fermions one must understand how continuum current algebra relations are realized on the lattice. This was explained in Refs. [5] and [6], and we recall here the relevant results.

In the continuum, the PCAC relation is

$$\langle\alpha|\partial_\mu A_\mu^a|\beta\rangle = \langle\alpha|\overline{\psi}\{\frac{\lambda^a}{2}, m_q\}\gamma_5\psi|\beta\rangle, \tag{7.1}$$

where $|\alpha\rangle$ and $|\beta\rangle$ are physical states, and $\lambda^a$ is a flavor Gell-Mann matrix. Individually $m_q$ and $\overline{\psi}\gamma_5\psi$ are scale dependent, and one must choose a particular scheme to precisely define them. The standard choice is the $\overline{\mathrm{MS}}$ scheme at a scale $\mu = 1$ GeV [20].

The lattice relation corresponding to Eq. (7.1) is [6]

$$\langle\alpha|Z_A\partial_\mu A_\mu^a|\beta\rangle = Z_P\langle\alpha|\overline{\psi}\{\frac{\lambda^a}{2}, m_q\}\gamma_5\psi|\beta\rangle + O(a), \tag{7.2}$$

where $m_q$, the quark mass, is related to the bare lattice quark mass by

$$m_q = Z_{\mathrm{mass}} m_q^{\mathrm{latt}}, \qquad m_q^{\mathrm{latt}} \equiv \frac{1}{2}\Big(\frac{1}{\kappa} - \frac{1}{\kappa_c}\Big). \tag{7.3}$$

In this and the following equations we assume degenerate quarks. We stress that $Z_{\mathrm{mass}}$ is the mass renormalization and differs from the finite constant $Z_m$ defined in Ref. [6] ($Z_m = Z_P Z_{\mathrm{mass}}$ in our notation). In Ref. [21] it was shown that for $m_q \to 0$, $Z_{\mathrm{mass}} = Z_S^{-1}$. In Appendix A we rederive this result in perturbation theory to all orders, and argue that it remains true for general $m_q$ in the continuum limit. For this reason, we will express all our results in terms of $Z_S$ and $Z_P$ by requiring $Z_{\mathrm{mass}} = Z_S^{-1}$ identically. It should be borne in mind, however, that for non-zero lattice spacing there are $O(ma)$ corrections to the individual $Z$ factors which depend on the initial and final states. Thus $Z_S$ factors extracted from different matrix elements may differ by terms of $O(ma)$.

For Wilson fermions, the absence of chiral symmetry means that $Z_S$ and $Z_P$ differ, and as discussed in the introduction they must either be calculated perturbatively or else fixed by some phenomenological requirement, since the continuum quantities we are comparing to are defined perturbatively. Their ratio $Z_P/Z_S$, however, is finite and can be extracted non-perturbatively, as we now describe.

From Eq. (7.2) we can extract $m_q Z_P/Z_A$ . To do this we calculate the ratios

$$\begin{aligned} -\frac{M_\pi}{2}\,\frac{\langle 0|A_4(t)J(0)|0\rangle}{\langle 0|P(t)J(0)|0\rangle} &\overset{t\to\infty}{\sim} \frac{Z_P}{Z_A}\, m_q^a \\ \frac{1}{2}\,\frac{\langle 0|\partial_4 A_4(t)J(0)|0\rangle}{\langle 0|P(t)J(0)|0\rangle} &\overset{t\to\infty}{\sim} \frac{Z_P}{Z_A}\, m_q^c, \end{aligned} \tag{7.4}$$

where $P$ and $A_4$ are the local operators and the smeared source $J = \pi$. Thus, given $Z_A$ and assuming $m_q = Z_S^{-1} m_q^{\mathrm{latt}}$, we can make a non-perturbative evaluation of $Z_P/Z_S$. Two other estimates, $m_q^b$ and $m_q^d$, are obtained by substituting $J = \pi_2$. All four methods give consistent results, as shown in Table 8. The quality of the signal is displayed in Figs. 7a and 7b. Also given in Table 8 are the values for $m_q^{\mathrm{latt}}$ and the results for $Z_P/Z_S$. Our values are consistent with the earlier estimate of Ref. [7], $Z_P/Z_S \approx 0.7$.

To calculate the chiral condensate we use two variants of the method suggested in Ref. [6]. This is based upon the continuum Ward Identity

$$\langle\overline{\psi}\psi\rangle^{\mathrm{WI}} \equiv \langle 0|S(0)|0\rangle = \lim_{m_q\to 0} m_q \int d^4x \langle 0|P(x)P(0)|0\rangle, \tag{7.5}$$

where $\langle\overline{\psi}\psi\rangle^{\mathrm{WI}}$ is the chiral condensate per light flavor. The lattice equivalent of this is [6]

$$\langle\overline{\psi}\psi\rangle^{\mathrm{WI}} = \lim_{m_q\to 0} m_q \sum_x \langle 0|Z_P P(x) Z_P P(0)|0\rangle. \tag{7.6}$$

Using Eq. (7.4) one can rewrite this as

$$\langle\overline{\psi}\psi\rangle^{\mathrm{WI}} = \lim_{m_q\to 0} \frac{Z_A}{2Z_P}\,\frac{\langle 0|\partial_4 A_4(t)J(0)|0\rangle}{\langle 0|P(t)J(0)|0\rangle} Z_P^2 \sum_{t'} \langle 0|P(t')P(0)|0\rangle, \tag{7.7}$$

where the ratio of correlators is evaluated at large $t$. We cannot use (7.7) since we only have $LS$ and $SS$ correlators available, so instead we use

$$\frac{\langle\overline{\psi}\psi\rangle^{\widetilde{\mathrm{WI}}}}{Z_P Z_A} = \lim_{m_q\to 0} \frac{1}{2}\,\frac{\langle 0|\partial_4 A_4(t)J(0)|0\rangle}{\langle 0|P^{\mathrm{smeared}}(t)J(0)|0\rangle} \sum_{t'} \langle 0|P(t')P^{\mathrm{smeared}}(0)|0\rangle. \tag{7.8}$$

This is equivalent to Eq. (7.7) if the pion pole dominates the sum on the right hand side, which occurs in the limit $m_q \to 0$. The results are given in Table 9 for the two choices of the source $J$. A typical example of the quality of the signal is shown in Fig. 8a for $J = \pi$.

A variant of this method is to assume pion dominance of Eq. (7.7), and derive a lattice version of the Gell-Mann, Oakes, Renner relation [22] [20]

$$\langle\overline{\psi}\psi\rangle^{\rm GMOR} = \lim_{m_q\to 0} -\frac{f_\pi^2 M_\pi^2}{4m_q}. \tag{7.9}$$

Using Eqs. (5.2) and (7.4), this can be extracted from the combination of correlators

$$\frac{\langle 0|P(t)J(0)|0\rangle\langle 0|A_4(t)J(0)|0\rangle}{\langle 0|J(t)J(0)|0\rangle} \quad \stackrel{t\to\infty}{\sim} \quad \frac{\langle\overline{\psi}\psi\rangle^{\rm GMOR}}{Z_P Z_A} e^{-M_\pi t}, \tag{7.10}$$

where the pion source can again be $J = \pi$ or $\pi_2$. (To derive this result one must bear in mind the normalization factors of $2M_\pi$, as shown explicitly in Eq. (5.2).) The results using this method are also given in Table 9, and in Fig. 8b we show a typical $m_{\rm eff}(t)$ plot for the above ratio.

These two methods should only agree for $m_q = 0$, in which limit they give the condensate. In an expansion

$$\langle\overline{\psi}\psi\rangle = \langle\overline{\psi}\psi\rangle_{m_q=0} + c m_q + \ldots, \tag{7.11}$$

the linear (and higher order) terms are contaminated by lattice artifacts. We also include in the Table 9 the results of linearly extrapolating $\langle\overline{\psi}\psi\rangle$ to $\kappa_c = 0.15704$. It is very encouraging to see that the four values agree with each other within our statistical resolution, and this leads us to believe that the results are physical. Therefore we compare them with those obtained using staggered fermions and with the experimental values in Sec. 9.

## 8. Second moment of the quark distribution amplitude in the pion

The quark distribution amplitude of a hadron is a wave function describing the distribution of the hadron momentum between valence quarks near the light cone limit. For the pion, the second moment of the distribution amplitude, $\langle\xi^2\rangle$, parameterizes the matrix element of an axial operator with two derivatives:

$$\begin{aligned} \langle 0|A_{\mu\nu\rho}(0)|\pi(\vec{p})\rangle &= Z^{-1} f_\pi \langle\xi^2\rangle p_\mu p_\nu p_\rho, \\ A_{\mu\nu\rho}(x) &= (-i)^2\overline{\psi}(x)\gamma_5\gamma_\mu \overleftrightarrow{D}_\nu \overleftrightarrow{D}_\rho \psi(x) - \text{traces}. \end{aligned} \tag{8.1}$$

Eqn. (8.1) holds independent of the symmetrization of Lorentz indices, though for certain analyses such a symmetrization is desirable in order to project out an operator of definite twist. On the lattice, the renormalization constant $Z$ is unknown, and depends on the Lorentz indices. In one loop perturbation theory, assuming that the dominant contribution to the renormalization comes from the tadpole terms which are independent of the Lorentz indices, one can estimate $Z \sim 1.3$. There is no reason to trust this perturbative result, however, and the results given below support a value substantially larger.

Using the lattice transcription of operators given in Ref. [23], we have measured $\langle\xi^2\rangle^{\rm latt} \equiv Z^{-1}\langle\xi^2\rangle$ from the following correlator ratios:

$$\begin{aligned} R_J^{[433]} &= \frac{\langle 0|A_{[433]}(\vec{p},t)J(0)|0\rangle}{\langle 0|A_4(\vec{p},t)J(0)|0\rangle} = \langle\xi^2\rangle^{\rm latt}\ p_3 p_3, \\ R_J^{[343]} &= \frac{\langle 0|A_{[343]}(\vec{p},t)J(0)|0\rangle}{\langle 0|A_4(\vec{p},t)J(0)|0\rangle} = \langle\xi^2\rangle^{\rm latt}\ p_3 p_3, \\ R_J^{(433)} &= \frac{\langle 0|A_{(433)}(\vec{p},t)J(0)|0\rangle}{\langle 0|A_4(\vec{p},t)J(0)|0\rangle} = \langle\xi^2\rangle^{\rm latt}\ p_3 p_3, \end{aligned} \tag{8.2}$$

where $J = \pi, \pi_2$ and $\vec{p} = (0, 0, 2\pi/16)$. Square brackets around indices indicate that an appropriate combination of operators has been taken to non-perturbatively subtract quadratic divergences [24] and round brackets indicate that symmetrization over the Lorentz indices has also been performed:

$$\begin{aligned} A_{[433]} &= A_{433} - A_{411}, \\ A_{[343]} &= A_{343} - A_{141}, \\ A_{(433)} &= \big(A_{[433]} + 2A_{[343]}\big)/3. \end{aligned} \tag{8.3}$$

Our results are given in Table 10. Examples of the quality of the data are shown in Figs. 9a and 9b. In general using $\pi$ as the pion source seems to give better results: with $\pi_2$ it takes longer to reach the asymptotic plateau.

For the symmetrized operator, $A_{(433)}$, the results are consistent with a previous analysis [23] in which configurations with dynamical quarks with masses in the range $m_s < m_q < 3m_s$ were used. This suggests that $\langle\xi^2\rangle^{\rm latt} \approx 0.1$. However, examination of the two unsymmetrized operators separately reveals a surprising feature: $\langle\xi^2\rangle^{\rm latt}$ from the operator with a time derivative, $A_{[343]}$, is much larger than from that with only spatial derivatives, $A_{[433]}$.

There are various possible explanations of this discrepancy, but note in particular that $A_{[433]}$ and $A_{[343]}$ renormalize differently (they do, of course, mix). This can resolve the problem only if there are large non-perturbative contributions to the renormalization constants. Lattice spacing corrections may also play a large role, and repeating these calculations with an improved fermion action should be very helpful.

We have also tried to determine $\langle\xi^2\rangle^{\rm latt}$ from $A_{423}$ at $\vec{p} = (0, 2\pi/16, 2\pi/16)$. The results are given in Table 11, though we are less confident of the signal at this higher momentum. We associate the poor signal with the lack of a plateau in the effective mass plot for the pion correlator with two units of momentum. The renormalization of $A_{423}$ is independent of the ordering of the indices, and the data do not show any significant difference.

## 9. Lattice scale and comparison of results

In order to study the chiral limit, we extrapolate our results for the various quantities to the value of $\kappa$ at which the pion mass vanishes. This is shown in Fig. 10, where we have made separate extrapolation for baryon masses obtained using Wuppertal (superscript $a$) and wall (superscript $b$) sources. We extrapolate using $m_q^{\rm latt}$. We find $\kappa_c = 0.15704$, in agreement with the result given in

Ref. [12]. The results of these fits are:

$$\begin{aligned} M_\pi^2 a^2 &= 0.000(16) + 2.1(3)\ m \\ f_\pi a &= 0.070(14) + 0.38(27)\ m \\ M_\rho a &= 0.313(33) + 2.3(6)\ m \\ M_N^a a &= 0.521(53) + 3.4(1.0)\ m \\ M_N^b a &= 0.486(91) + 3.5(1.6)\ m \\ M_\Delta^a a &= 0.700(72) + 1.9(1.4)\ m \\ M_\Delta^b a &= 0.58(13) + 2.9(2.1)\ m \end{aligned} \tag{9.1}$$

from which we can extract the lattice scale. Using the experimental values for $f_\pi$, $M_\rho$, $M_N$ and $M_\Delta$ we get:

$$\begin{aligned} a^{-1}(f_\pi) &= 1.9(4)\ \mathrm{GeV} \\ a^{-1}(M_\rho) &= 2.5(3)\ \mathrm{GeV} \\ a^{-1}(M_N^a) &= 1.8(2)\ \mathrm{GeV} \\ a^{-1}(M_N^b) &= 1.93(36)\ \mathrm{GeV} \\ a^{-1}(M_\Delta^a) &= 1.76(18)\ \mathrm{GeV} \\ a^{-1}(M_\Delta^b) &= 2.1(5)\ \mathrm{GeV}. \end{aligned} \tag{9.2}$$

The large errors in the extrapolated values reflect the fact that we have data at just two values of $\kappa$. The $a^{-1}$ extracted from $M_\rho$ is significantly larger. Also, the wall source estimates for the nucleon mass give a slightly better value for the ratio $M_N/M_\rho$.

A linear extrapolation of $\langle\overline{\psi}\psi\rangle$ to the chiral limit gives,

$$\begin{aligned} \langle\overline{\psi}\psi\rangle_{m_q=0} &= -Z_P Z_A\ 0.0058(18)\ a^{-3} \\ &\sim -Z_P Z_A\ 0.034\ \mathrm{GeV}^3, \end{aligned} \tag{9.3}$$

where we have used $a^{-1} = 1.8$ GeV. Assuming $Z_P Z_A \sim 1$, we note that this result is approximately three times larger than the continuum value $\langle\overline{u}u\rangle = -0.0114\ \mathrm{GeV}^3$ [20]. To make a meaningful comparison we need to estimate $Z_P$.

The 1-loop perturbative result for $Z_P$ using Wilson fermions is discussed in Appendix A. It is

$$Z_P = 1 + \frac{g^2 C_F}{16\pi^2}\left(6\log(a\mu) - 22.596\right), \tag{9.4}$$

where $C_F = 4/3$. This estimate does not address possible $O(a)$ and other non-perturbative effects. We use the continuum $\overline{\rm MS}$ renormalization scheme at $\mu = 1$ GeV so that $Z_P = 1 - 0.221 g^2$ for $a^{-1} = 1.8$ GeV. This leaves one source of uncertainty in applying this formula, namely the choice of $g^2$. Using $g^2 \approx 2$, as advocated by Lepage and Mackenzie [25], we get $Z_A Z_P \approx 0.5$. Given the many uncertainties in this estimate, what we can say with some confidence is that including these corrections moves the lattice estimate closer to the continuum result.

Using Eq. (9.1) we can also attempt to extract the mass of the strange quark. Our estimate varies by a factor of two depending on the method we use. For example, by demanding that the ratio $2M_K^2/f_\pi^2$ attain its physical value at $m = m_s$ we get $m_s^{\rm latt} a \approx 0.066$, while the ratio $M_{K^*}^2/2M_K^2$ gives $m_s^{\rm latt} a \approx 0.036$. Translated into physical units these correspond to $Z_S m_s = 81$ MeV and 44 MeV respectively. If we assume $Z_S \approx 1$, then these values are a factor of 2–3 smaller than the conventional estimate. This discrepancy is similar to the results obtained with staggered fermions [13]. Once again, a value of $Z_S < 1$ would increase these estimates.

The combination $m_s \langle \overline{\psi}\psi \rangle_{m_q=0}$ can be extracted with less ambiguity because it involves the finite ratio $Z_P/Z_S$, for which we have a non-perturbative estimate. Taking $Z_A = 0.86$ and $Z_P/Z_S = 0.68$, and using the two extreme values for $m_s$ quoted above, we find $-0.0024$ GeV$^4$ and $-0.0013$ GeV$^4$, to be compared to the experimental value of approximately $-0.0017$ GeV$^4$.

## 10. Conclusions

We show that both Wuppertal and wall quark sources yield very good signals for the $\pi$ and $\rho$ mesons at $\vec{p} = (0,0,0)$. There is an unambiguous plateau in the effective mass plots for all three types of correlators studied and the mass estimates are consistent. Since the estimates converge from opposite directions with the two kinds of sources, consistency of the results implies that we have extracted the asymptotic value. We find that the signal in the Wuppertal source pion correlator with $\vec{p} = (0,0,2\pi/16)$ is good enough to allow the calculation of matrix elements with non-zero momentum flow.

There are strong indications that we have not extracted the asymptotic value of the baryon masses. The disturbing difference between the results from the Wuppertal and wall sources needs to be understood before we can quote the baryon masses with confidence.

The signal in the ratios of Wuppertal source correlators used to extract $f_\pi$, $f_V^{-1}$, $m_q$ and $\langle \overline{\psi}\psi \rangle$ is very good. We show how to extract the current quark mass and the chiral condensate using smeared sources, although comparing with experiment is difficult in the absence of reliable values for $Z_P$ and $Z_S$. We show internal consistency of the results by using different hadronic operators and by using different combinations of correlators. We find that our estimate of $f_V^{-1}$ is in good agreement with experimental values.

Lastly, we have calculated the second moment of the pion distribution amplitude, $\langle \xi^2 \rangle^{\rm latt}$. We find a significant difference between the results extracted from $A_{[433]}$ and $A_{[343]}$ at $\vec{p} = (0,0,2\pi/16)$. The average value of $\langle \xi^2 \rangle^{\rm latt} \sim 0.1$ is consistent with earlier results obtained using lattices with two dynamical flavors of Wilson fermions.

**Acknowledgements**

The $16^3 \times 40$ lattices were generated at NERSC at Livermore using a DOE allocation. The calculation of quark propagators and the analysis has been done at the Pittsburgh Supercomputing Center, San Diego Supercomputer Center, NERSC and Los Alamos National Laboratory. We are very grateful to Jeff Mandula, Norm Morse, Ralph Roskies, Charlie Slocomb and Andy White for their support of this project. We thank C. Bernard and G. Martinelli for discussions. SRS is supported in by the DOE contract DE-FG09-91ER40614 and an Alfred P. Sloan Fellowship. This research was supported in part by the National Science Foundation under Grant No. PHY89-04035.

## Appendix A. Perturbative evaluation of $Z_S$, $Z_P$ and $Z_{\rm mass}$

The perturbative calculations required to evaluate the constants $Z_S$, $Z_P$ and $Z_{\rm mass}$ with Wilson fermions have been performed by a variety of authors [26] [27] [28] [29] [30] [31]. It is, however, rather difficult to extract the precise values from these papers, and for this reason we collect some of the relevant results here. We restrict ourselves to $r=1$, and take $\overline{\rm MS}$ as our continuum scheme.

The calculation of the renormalization constants requires the evaluation of the fermion self-energy (wave function and mass renormalization) and vertex corrections. The continuum fermion self-energy in the $\overline{\rm MS}$ scheme has the form

$$\Sigma^{\overline{\rm MS}}(p,m) = i\gamma\cdot p\,\Sigma_1^{\overline{\rm MS}}(p,m) + m\,\Sigma_2^{\overline{\rm MS}}(p,m), \tag{A.1}$$

with the 1-loop results

$$\begin{aligned}
\Sigma_1^{\overline{\rm MS}}(p,m) &= \lambda\Big\{1 + 2\int_0^1 dx\,(1-x)\log\Big(\frac{xm^2+x(1-x)p^2}{\mu^2}\Big)\Big\},\\
\Sigma_2^{\overline{\rm MS}}(p,m) &= \lambda\Big\{2 + 4\int_0^1 dx\,\log\Big(\frac{xm^2+x(1-x)p^2}{\mu^2}\Big)\Big\},
\end{aligned} \tag{A.2}$$

where $\lambda = g^2C_F/(16\pi^2)$ and $C_F = 4/3$. The lattice self-energy is

$$\Sigma^{\rm latt}(p,m) = \frac{1}{a}\,\Sigma_0^{\rm latt} + i\gamma\cdot p\,\Sigma_1^{\rm latt}(p,m) + m\,\Sigma_2^{\rm latt}(p,m), \tag{A.3}$$

where at 1-loop

$$\begin{aligned}
\Sigma_0^{\rm latt} &= \lambda(-51.435),\\
\Sigma_1^{\rm latt}(p,m) &= \lambda\Big\{13.852 + 2\int_0^1 dx\,(1-x)\log\big(a^2(xm^2+x(1-x)p^2)\big)\Big\}\\
&\qquad + O(ma,pa),\\
\Sigma_2^{\rm latt}(p,m) &= \lambda\Big\{1.901 + 4\int_0^1 dx\,\log\big(a^2(xm^2+x(1-x)p^2)\big)\Big\}\\
&\qquad + O(ma,pa).
\end{aligned} \tag{A.4}$$

We have taken the most precise values available for the finite constants [29]. These are obtained by numerical integration and are accurate to better than one part in the last decimal place. The linearly divergent piece, $\Sigma_0^{\rm latt}$, shifts the position of $\kappa_c$, and plays no role in the following discussion. Neglecting this term, and adopting a slightly modified version of the notation of Ref. [27], we write the difference between continuum and lattice schemes (in the limit $a\to 0$) as

$$\Sigma^{\overline{\rm MS}}(p,m) - \Sigma^{\rm latt}(p,m) = i\gamma\cdot p\,\Delta_{\Sigma_1} + m\,\Delta_{\Sigma_2}. \tag{A.5}$$

At 1-loop, let us define $\Delta_i = \lambda\Delta_i^{(1)}$. Then

$$\Delta_{\Sigma_1}^{(1)} = -2\log(a\mu) - 12.852, \qquad \Delta_{\Sigma_2}^{(1)} = -8\log(a\mu) + 0.099. \tag{A.6}$$

With these results the 1-loop mass renormalization constant is

$$\begin{aligned}
Z_{\rm mass} &= 1 + \lambda(\Delta_{\Sigma_2}^{(1)} - \Delta_{\Sigma_1}^{(1)}),\\
&= 1 + \lambda\big(-6\log(a\mu) + 12.951\big).
\end{aligned} \tag{A.7}$$

In Ref. [26] the result for $Z_{\rm mass}$ is in error, because the expression for $\Delta_{\Sigma_2}^{(1)}$ is incorrect. This error was pointed out in Ref. [28], where a result consistent with Eq. (A.7) is obtained.

Notice that like the $\Delta$'s, the perturbative $Z$'s are defined neglecting terms of $O(a)$ and, in particular, terms of $O(ma)$.

To calculate $Z_S$ and $Z_P$ one also needs the difference between the $\overline{\rm MS}$ and lattice vertex corrections for insertions of $\overline{\psi}\psi$ and $\overline{\psi}\gamma_5\psi$. Following Ref. [27], we write these differences as $\Delta_1$ and $\Delta_{\gamma_5}$ respectively. The above mentioned error in Ref. [26] propagates into an error in the 1-loop result for $\Delta_1 = \lambda\Delta_1^{(1)}$ given in Ref. [27]. The result for $\Delta_{\gamma_5}^{(1)}$ is correct. More precise 1-loop values may be deduced by combining the results of Refs. [31], [30] and [27]:

$$\Delta_1^{(1)} = +8\log(a\mu) - 0.100, \qquad \Delta_{\gamma_5}^{(1)} = +8\log(a\mu) - 9.744. \tag{A.8}$$

To extract these results one needs to know that in the dimensional reduction scheme used in Ref. [30], both $\Delta_1^{(1)}$ and $\Delta_{\gamma_5}^{(1)}$ are larger by +2 than in the $\overline{\rm MS}$ scheme.

With these results at hand we can now calculate the 1-loop renormalization constants

$$
\begin{aligned}
Z_S &= 1 + \lambda(\Delta_1^{(1)} + \Delta_{\Sigma_1}^{(1)}), \\
&= 1 + \lambda\big( + 6\log(a\mu) - 12.952\big); \\
Z_P &= 1 + \lambda(\Delta_{\gamma_5}^{(1)} + \Delta_{\Sigma_1}^{(1)}), \\
&= 1 + \lambda\big( + 6\log(a\mu) - 22.596\big).
\end{aligned}
\tag{A.9}
$$

Eqs. (A.7) and (A.9) imply that, because $\Delta_1^{(1)} = -\Delta_{\Sigma_2}^{(1)}$ within the numerical errors, the relation $Z_{\rm mass} Z_S = 1$ is satisfied. As mentioned in the text, this relation is expected on the basis of the functional integral derivation given in Ref. [21]. In fact, as we now discuss, it is straightforward to understand this result in perturbation theory.

To establish the equality $\Delta_1 = -\Delta_{\Sigma_2}$ at 1-loop we begin by noting that to this order

$$
\lambda \Delta_{\Sigma_2}^{(1)} = \frac{\partial}{\partial m} \left( \Sigma^{\overline{\rm MS}}(p,m) - \Sigma^{\rm latt}(p,m) \right) \Big|_{m=0} \ , \tag{A.10}
$$

which follows from Eqs. (A.5) and (A.6). The crucial observation is that the derivative with respect to $m$ inserts minus the scalar density on the internal fermion line, and so gives the difference of continuum and lattice vertex diagrams. These are precisely the diagrams which, in the limit $m \to 0$, define $-\Delta_1^{(1)}$. Note that terms suppressed by powers of $ma$ (which are present in $\Sigma^{\rm latt}$) are irrelevant since the $\Delta$'s are defined dropping all terms of $O(a)$.

This derivation extends straighforwardly to all orders. Taking the derivative with respect to $m$ in a general self energy diagram is equivalent to the insertion of minus the scalar density in all possible ways, and so by definition gives the corresponding set of graphs for $-\Delta_1$. Note that this is true for any $r$. The crucial assumption is that the difference of continuum and lattice self-energies which defines the $\Delta$'s is infra-red finite, and thus has a smooth chiral limit. This forbids contributions to the difference of $\Sigma$'s which would make the derivative infrared singular, i.e. terms of the form $\log(ma)$, $\log(m^2/p^2)$, $p^2/m^2$, etc. Dimensional analysis then shows that the only dependence on $m$ comes from polynomials in $ma$ and $ma\log(ma)$, which give no contribution in the continuum limit.

This discussion makes clear that the relation $Z_S Z_{\rm mass} = 1$, which we use in the text, holds exactly in perturbation theory. The individual renormalization factors we observe in a simulation at finite lattice spacing, however, also contain $O(ma)$ corrections, the precise values of which depend upon the external states. Consequently, the replacement $Z_{\rm mass} \to Z_S^{-1}$ may have corrections of $O(ma)$ if the two are extracted from different matrix elements.

Finally, we give the finite constant

$$
\begin{aligned}
Z_{\rm mass} Z_P &= 1 + \lambda(\Delta_{\gamma_5}^{(1)} + \Delta_{\Sigma_2}^{(1)}), \\
&= 1 + \lambda(-9.644),
\end{aligned}
\tag{A.11}
$$

which is denoted $Z_m$ in Ref. [6]. Note that in order to reproduce our non-perturbative result using this formula, one would have to take $g^2 \approx 4$.

| $\Gamma$ | $\pi$ | $\pi_2$ | $\rho$ | $\rho_2$ | $a_1$ | $a_0$ | $b_1$ |
|---|---|---|---|---|---|---|---|
| $SS$ | 0.53 | 1.25 | 1.23 | 0.69 | 1.42 | 0.82 | 1.02 |
| | $10-22$ | $8-24$ | $10-20$ | $8-16$ | $5-13$ | $5-10$ | $5-10$ |
| | 0.362(6) | 0.368(10) | 0.459(14) | 0.465(11) | 0.759(31) | 0.708(37) | 0.772(38) |
| $LS$ | 0.94 | 0.89 | 0.48 | 1.16 | 0.36 | 1.27 | 0.62 |
| | $10-25$ | $15-25$ | $10-22$ | $10-19$ | $7-11$ | $7-11$ | $7-11$ |
| | 0.365(6) | 0.364(14) | 0.460(7) | 0.467(8) | 0.732(38) | 0.717(40) | 0.766(44) |
| $LW$ | 0.63 | 0.83 | 0.35 | 0.25 | 1.13 | 0.57 | 0.42 |
| | $10-22$ | $9-30$ | $10-20$ | $8-16$ | $6-10$ | $6-10$ | $6-10$ |
| | 0.361(4) | 0.365(6) | 0.463(6) | 0.460(9) | 0.727(30) | 0.676(34) | 0.714(31) |

**Table 1.** Meson masses at $\kappa = 0.154$. For each operator and correlator we give the $\chi^2$ per degree of freedom, the fit range and the mass estimate.

| $\Gamma$ | $N^+$ | $N^-$ | $\Delta_1^+$ | $\Delta_2^+$ | $\Delta_1^-$ | $\Delta_2^-$ |
|---|---|---|---|---|---|---|
| $SS$ | 0.60 | 1.58 | 0.65 | 1.55 | 1.51 | 0.48 |
| | $8-16$ | $6-10$ | $9-16$ | $8-15$ | $5-12$ | $5-10$ |
| | 0.733(17) | 1.034(48) | 0.796(47) | 0.824(24) | 1.105(47) | 1.134(36) |
| $LS$ | 0.86 | 0.47 | 0.40 | 0.68 | 1.24 | 0.46 |
| | $10-17$ | $7-10$ | $10-16$ | $10-16$ | $7-11$ | $7-11$ |
| | 0.740(14) | 1.069(42) | 0.805(14) | 0.822(12) | 1.111(69) | 1.130(37) |
| $LW$ | 0.30 | 0.40 | 0.28 | 0.25 | 1.87 | 0.73 |
| | $9-16$ | $6-10$ | $9-15$ | $9-15$ | $5-12$ | $6-10$ |
| | 0.708(18) | 0.935(23) | 0.767(21) | 0.761(20) | 0.958(25) | 1.009(42) |

**Table 2.** Baryon masses at $\kappa = 0.154$. For each operator and correlator we give the $\chi^2$ per degree of freedom, the fit range and the mass estimate.

| $\Gamma$ | $\pi$ | $\pi_2$ | $\rho$ | $\rho_2$ | $a_1$ | $a_0$ | $b_1$ |
|---|---|---|---|---|---|---|---|
| | 1.39 | 0.79 | 1.65 | 1.45 | 1.04 | 1.70 | 0.52 |
| $SS$ | $10-22$ | $7-16$ | $10-22$ | $8-20$ | $5-9$ | $4-8$ | $4-8$ |
| | 0.297(9) | 0.303(7) | 0.411(14) | 0.425(20) | 0.714(45) | 0.738(64) | 0.748(54) |
| | 1.60 | 1.15 | 0.44 | 1.40 | 0.28 | 1.75 | 0.93 |
| $LS$ | $10-25$ | $12-24$ | $10-22$ | $10-18$ | $7-10$ | $6-10$ | $7-10$ |
| | 0.297(9) | 0.298(17) | 0.411(10) | 0.428(13) | 0.700(43) | 0.696(44) | 0.744(55) |
| | 1.36 | 1.45 | 0.51 | 0.65 | 0.88 | 0.66 | 0.41 |
| $LW$ | $10-22$ | $8-20$ | $9-20$ | $8-18$ | $5-10$ | $5-10$ | $5-10$ |
| | 0.295(5) | 0.301(7) | 0.420(10) | 0.431(9) | 0.692(24) | 0.657(53) | 0.693(21) |

**Table 3.** Meson masses at $\kappa = 0.155$. For each operator and correlator we give the $\chi^2$ per degree of freedom, the fit range and the mass estimate.

| $\Gamma$ | $N^+$ | $N^-$ | $\Delta_1^+$ | $\Delta_2^+$ | $\Delta_1^-$ | $\Delta_2^-$ |
|---|---|---|---|---|---|---|
| | 0.70 | 1.21 | 0.94 | 2.35 | 1.81 | 0.48 |
| $SS$ | $8-17$ | $5-10$ | $8-17$ | $9-15$ | $6-12$ | $5-10$ |
| | 0.664(15) | 0.989(37) | 0.752(25) | 0.778(32) | 1.010(95) | 1.099(44) |
| | 0.80 | 1.67 | 0.82 | 1.01 | 0.76 | 0.18 |
| $LS$ | $9-17$ | $7-18$ | $10-16$ | $9-15$ | $7-11$ | $8-11$ |
| | 0.665(15) | 0.942(73) | 0.740(18) | 0.778(21) | 1.075(90) | 1.035(97) |
| | 0.35 | 0.38 | 0.93 | 0.63 | 0.57 | 0.52 |
| $LW$ | $8-16$ | $6-10$ | $8-14$ | $8-16$ | $4-8$ | $6-10$ |
| | 0.634(28) | 0.885(32) | 0.718(22) | 0.697(37) | 0.910(21) | 0.951(60) |

**Table 4.** Baryon masses at $\kappa = 0.155$. For each operator and correlator we give the $\chi^2$ per degree of freedom, the fit range and the mass estimate.

| $\kappa$ | $Lattice$ | $N_{conf}$ | $\pi$ | $\rho$ | $N$ | $\Delta$ | $f_\pi$ |
|---|---|---|---|---|---|---|---|
| 0.154 | $16^3 \times 40$ | 35 | 0.365(4) | 0.465(7) | 0.736(11) | 0.82(2) | 0.093(3) |
| 0.154 | $16^3 \times 80$ | $35^a$ | 0.364(6) | 0.460(7) | 0.737(14) | 0.82(2) | 0.091(4) |
| 0.154 | $16^3 \times 80$ | $35^b$ | 0.361(4) | 0.463(6) | 0.708(18) | 0.76(2) | |
| 0.154 | $18^3 \times 32$ | 104 | 0.361(1) | 0.463(3) | 0.721(7) | 0.782(10) | |
| 0.155 | $16^3 \times 40$ | 35 | 0.301(6) | 0.420(12) | 0.663(15) | 0.78(2) | 0.086(3) |
| 0.155 | $16^3 \times 80$ | $35^a$ | 0.297(9) | 0.411(10) | 0.665(15) | 0.78(2) | 0.087(4) |
| 0.155 | $16^3 \times 80$ | $35^b$ | 0.295(5) | 0.420(10) | 0.634(28) | 0.70(4) | |
| 0.155 | $18^3 \times 32$ | 104 | 0.295(1) | 0.422(4) | 0.651(10) | 0.723(14) | |
| 0.155 | $24^3 \times 32$ | 78 | 0.297(3) | 0.428(4) | 0.647(6) | 0.745(15) | 0.083(3) |

**Table 5.** Quenched Wilson Fermion Spectrum at $\beta = 6.0$. We present results from Wuppertal (denoted by superscript $a$ in column 3) and wall sources (superscript $b$) on $16^3 \times 80$ lattices separately. Our Wuppertal source results on both doubled and un-doubled lattices are the mean of the $LS$ and $SS$ values. We include results on $18^3 \times 32$ and on $24^3 \times 32$ lattices from the APE collaboration [12] for comparison. To get the value for $f_\pi$ in lattice units we have used $Z_A = 0.86$.

| | $f_\pi^a$ | $f_\pi^b$ | $f_\pi^c$ | $f_\pi^d$ |
|---|---|---|---|---|
| $\kappa = 0.1540$<br>$\vec{p} = (0,0,0)$ | 0.95<br>$15-22$<br>0.090(4) | 1.56<br>$12-22$<br>0.090(7) | 1.33<br>$12-22$<br>0.092(5) | 1.00<br>$12-22$<br>0.093(7) |
| $\kappa = 0.1540$<br>$\vec{p} = (0,0,1)$ | 0.73<br>$10-16$<br>0.093(12) | 1.84<br>$8-16$<br>0.094(5) | 0.21<br>$9-16$<br>0.094(7) | 0.32<br>$10-16$<br>0.094(8) |
| $\kappa = 0.1550$<br>$\vec{p} = (0,0,0)$ | 1.93<br>$8-24$<br>0.086(4) | 1.18<br>$10-20$<br>0.079(8) | 1.49<br>$8-28$<br>0.088(5) | 1.85<br>$9-21$<br>0.087(7) |
| $\kappa = 0.1550$<br>$\vec{p} = (0,0,1)$ | 0.92<br>$8-14$<br>0.090(7) | 1.23<br>$8-14$<br>0.085(4) | 0.93<br>$7-14$<br>0.084(9) | 0.23<br>$9-16$<br>0.088(8) |

**Table 6.** Results for the pseudoscalar decay constant, $f_\pi$, calculated in the four ways described in the text, and using $Z_A = 0.86$. For each measurement we give the $\chi^2$ per degree of freedom, the fit range and the estimate.

| $\kappa$ | $1/f_V^a$ | $1/f_V^b$ |
|---|---|---|
| 0.1540 | 1.03<br>8 – 18<br>0.24(2) | 1.53<br>10 – 20<br>0.25(2) |
| 0.1550 | 1.19<br>8 – 18<br>0.26(2) | 2.84<br>10 – 22<br>0.26(2) |

**Table 7.** The value of the vector meson decay constant, $f_V^{-1}$, calculated in the two ways described in the text, and using $Z_V = 0.57$. For each measurement we give the $\chi^2$ per degree of freedom, the fit range and the estimate.

| $\kappa$ | $Z_P m_q^a$ | $Z_P m_q^b$ | $Z_P m_q^c$ | $Z_P m_q^d$ | $m_q^{\mathrm{latt}}$ | $Z_P/Z_S$ |
|---|---|---|---|---|---|---|
| 0.1540 | 1.29<br>12 – 25<br>0.042(1) | 0.79<br>12 – 22<br>0.041(2) | 1.76<br>10 – 30<br>0.043(1) | 0.79<br>12 – 28<br>0.042(1) | 0.062 | 0.68(1) |
| 0.1550 | 1.41<br>12 – 20<br>0.027(1) | 0.85<br>12 – 22<br>0.027(2) | 1.52<br>12 – 28<br>0.028(1) | 1.06<br>12 – 28<br>0.028(1) | 0.042 | 0.66(2) |

**Table 8.** The value of $Z_P m_q$, calculated in the four ways described in the text, using $Z_A = 0.86$. For each measurement we give the $\chi^2$ per degree of freedom, the fit range and the estimate. Also given are the lattice quark mass and the renormalization constant $Z_P/Z_S$.

| $\kappa$ | $\langle\overline{\psi}\psi\rangle_{\pi}^{\widetilde{\mathrm{WI}}}$ | $\langle\overline{\psi}\psi\rangle_{\pi_2}^{\widetilde{\mathrm{WI}}}$ | $\langle\overline{\psi}\psi\rangle_{\pi}^{\mathrm{GMOR}}$ | $\langle\overline{\psi}\psi\rangle_{\pi_2}^{\mathrm{GMOR}}$ |
|---|---|---|---|---|
| 0.1540 | 1.12<br>12 − 22<br>−0.0080(8) | 0.84<br>8 − 18<br>−0.0079(8) | 1.78<br>10 − 20<br>−0.0147(8) | 1.22<br>8 − 18<br>−0.0149(8) |
| 0.1550 | 1.89<br>8 − 22<br>−0.0073(7) | 1.48<br>10 − 22<br>−0.0072(13) | 1.82<br>8 − 18<br>−0.0118(7) | 1.07<br>8 − 18<br>−0.0118(8) |
| 0.15704 | −0.0059(26) | −0.0058(42) | −0.0060(26) | −0.0056(29) |

**Table 9.** The value of the chiral condensate on the lattice. For each measurement we give the $\chi^2$ per degree of freedom, the fit range and the estimate. We have to multiply all numbers by $Z_P Z_A$ in order to get physical values. The result at $\kappa_c = 0.15704$ is obtained by linear extrapolation.

| $\kappa$ | $\langle\xi^2\rangle_{\pi}^{[433]}$ | $\langle\xi^2\rangle_{\pi}^{[343]}$ | $\langle\xi^2\rangle_{\pi}^{(433)}$ | $\langle\xi^2\rangle_{\pi_2}^{[433]}$ | $\langle\xi^2\rangle_{\pi_2}^{[343]}$ | $\langle\xi^2\rangle_{\pi_2}^{(433)}$ |
|---|---|---|---|---|---|---|
| 0.1540 | 0.61<br>5 − 10<br>0.06(2) | 0.51<br>5 − 10<br>0.13(2) | 0.82<br>5 − 10<br>0.10(2) | 0.47<br>10 − 16<br>0.07(4) | 1.27<br>10 − 20<br>0.16(3) | 1.32<br>6 − 16<br>0.11(1) |
| 0.1550 | 0.66<br>5 − 10<br>0.06(3) | 0.54<br>6 − 12<br>0.12(3) | 0.91<br>5 − 12<br>0.10(2) | 0.16<br>10 − 15<br>0.09(5) | 1.70<br>7 − 15<br>0.18(2) | 1.37<br>7 − 15<br>0.11(2) |

**Table 10.** The value of $\langle\xi^2\rangle^{\mathrm{latt}}$ from $A_{433}$ calculated in the six ways described in the text. For each measurement we give the $\chi^2$ per degree of freedom, the fit range used and the estimate.

| $\kappa$ | $\langle\xi^2\rangle_\pi^{423}$ | $\langle\xi^2\rangle_\pi^{243}$ | $\langle\xi^2\rangle_\pi^{(423)}$ | $\langle\xi^2\rangle_{\pi_2}^{423}$ | $\langle\xi^2\rangle_{\pi_2}^{243}$ | $\langle\xi^2\rangle_{\pi_2}^{(423)}$ |
|---|---|---|---|---|---|---|
| 0.1540 | 1.34<br>$6-12$<br>0.14(2) | 0.70<br>$6-12$<br>0.10(2) | 1.62<br>$6-15$<br>0.12(2) | 2.09<br>$8-12$<br>0.08(4) | 2.19<br>$8-15$<br>0.11(3) | 3.58<br>$7-14$<br>0.11(2) |
| 0.1550 | 0.82<br>$6-12$<br>0.15(4) | 1.17<br>$5-10$<br>0.13(3) | 1.05<br>$6-12$<br>0.12(3) | 1.77<br>$7-12$<br>0.08(4) | 2.02<br>$8-15$<br>0.09(3) | 3.66<br>$6-12$<br>0.15(3) |

**Table 11.** The value of $\langle\xi^2\rangle^{\rm latt}$ from $A_{423}$ calculated in the six ways described in the text. For each measurement we give the $\chi^2$ per degree of freedom, the fit range used and the estimate.

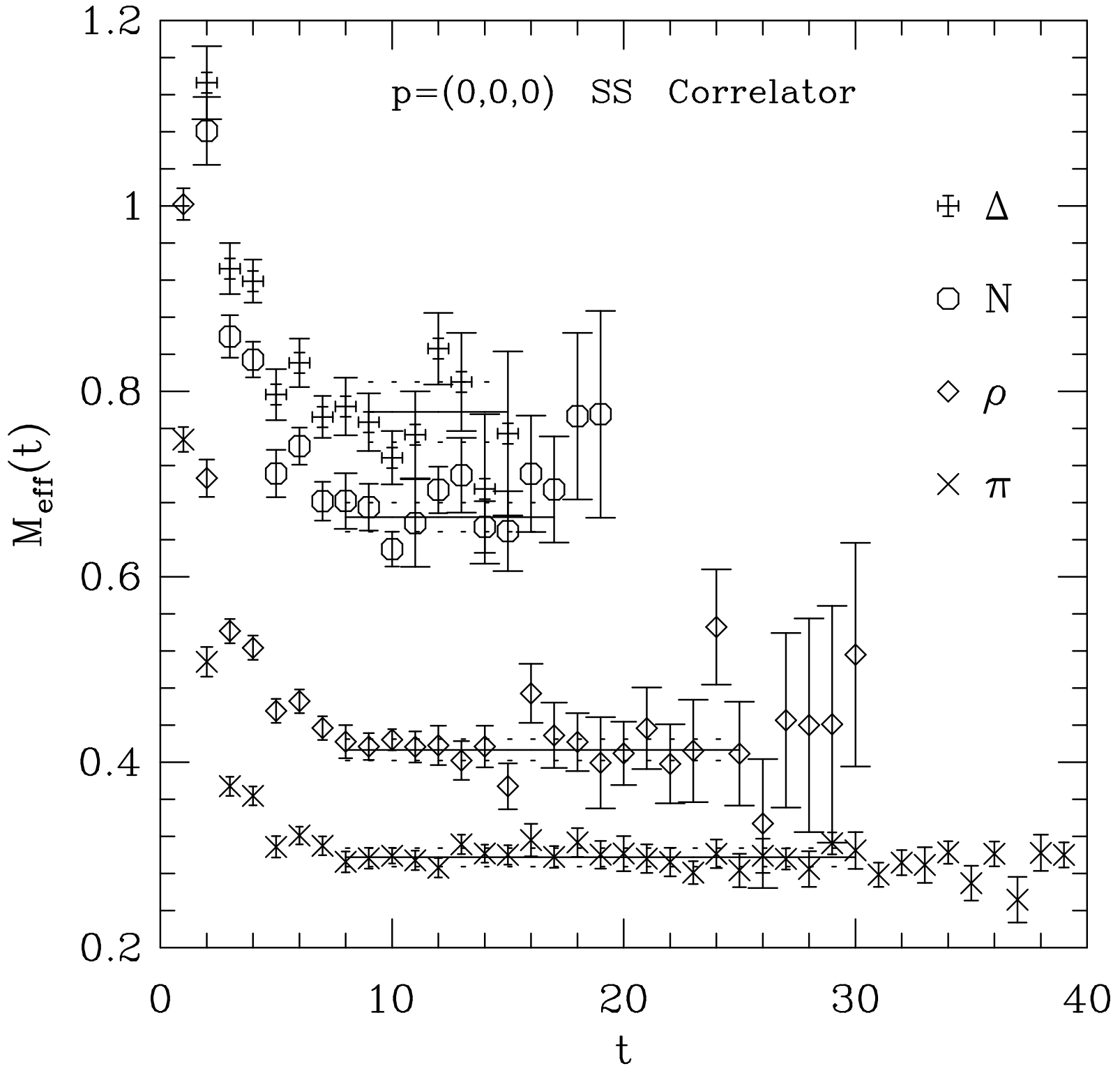

**Fig. 1a.** The effective mass plots for the $\pi$, $\rho$, nucleon and the $\Delta$ at $\kappa = 0.155$ using $SS$ correlators.

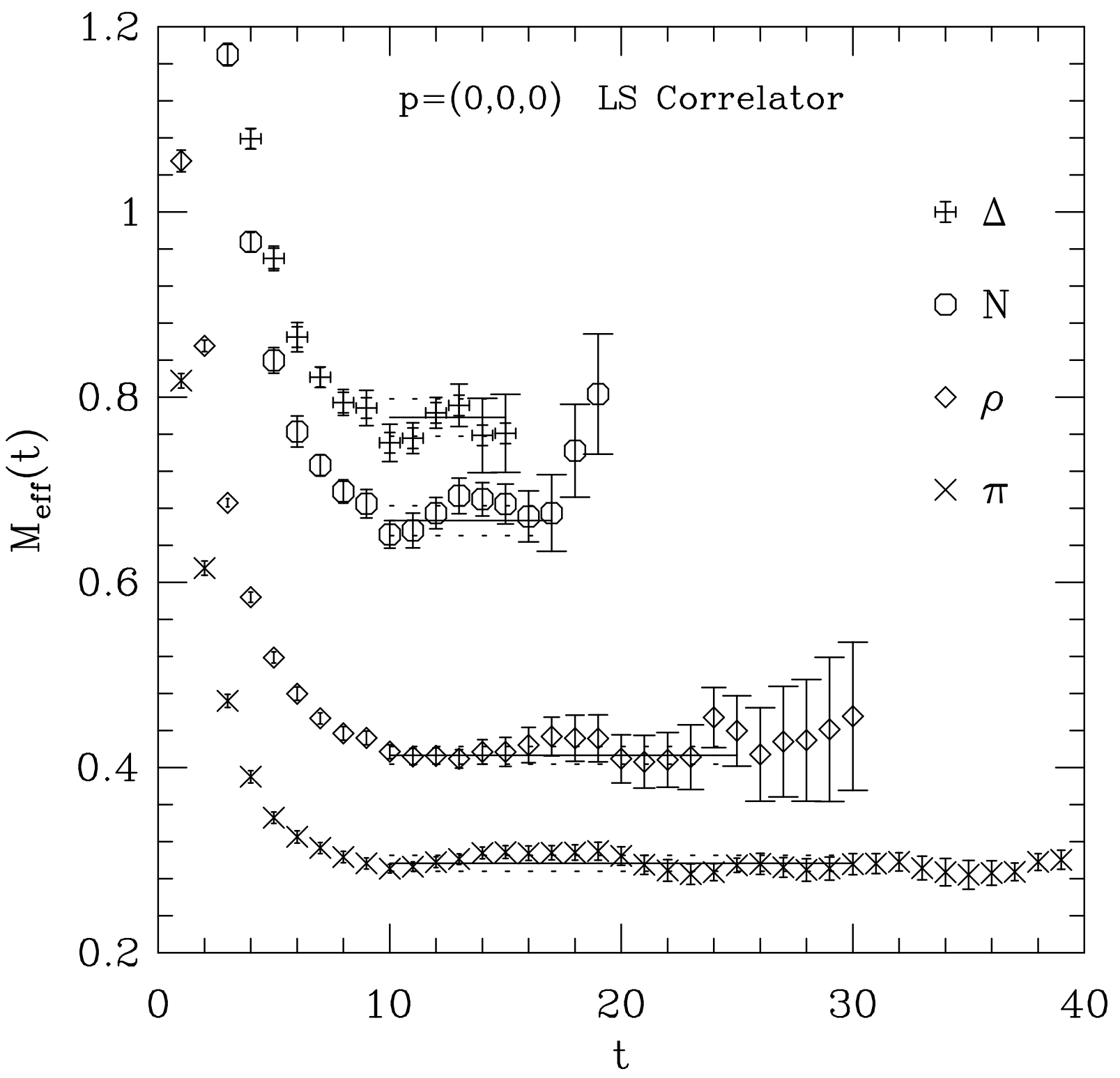

**Fig. 1b.** Same as in Fig. 1a except the data are for *LS* correlators,

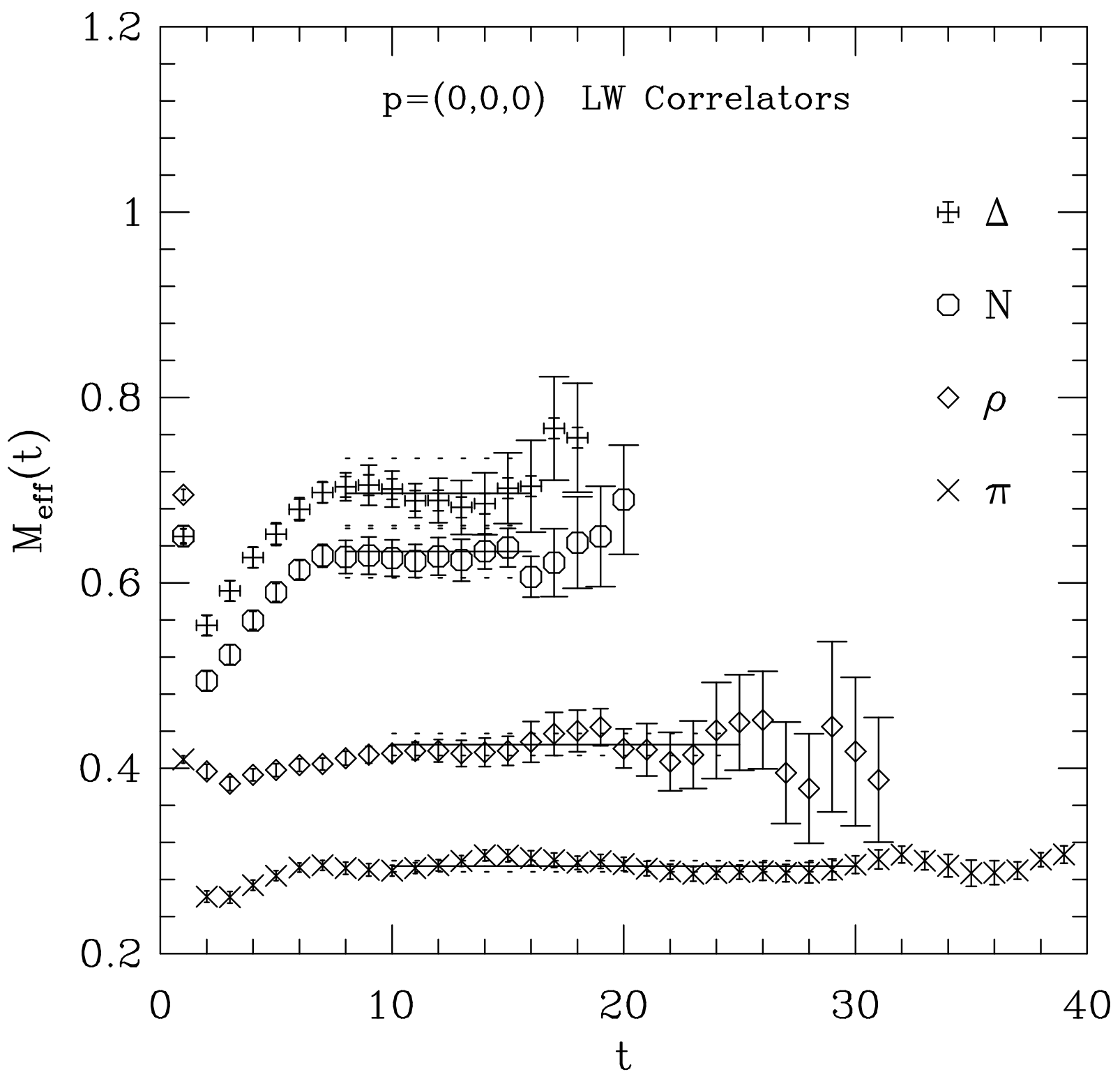

**Fig. 1c.** Same as in Fig. 1a except the data are for *LW* correlators.

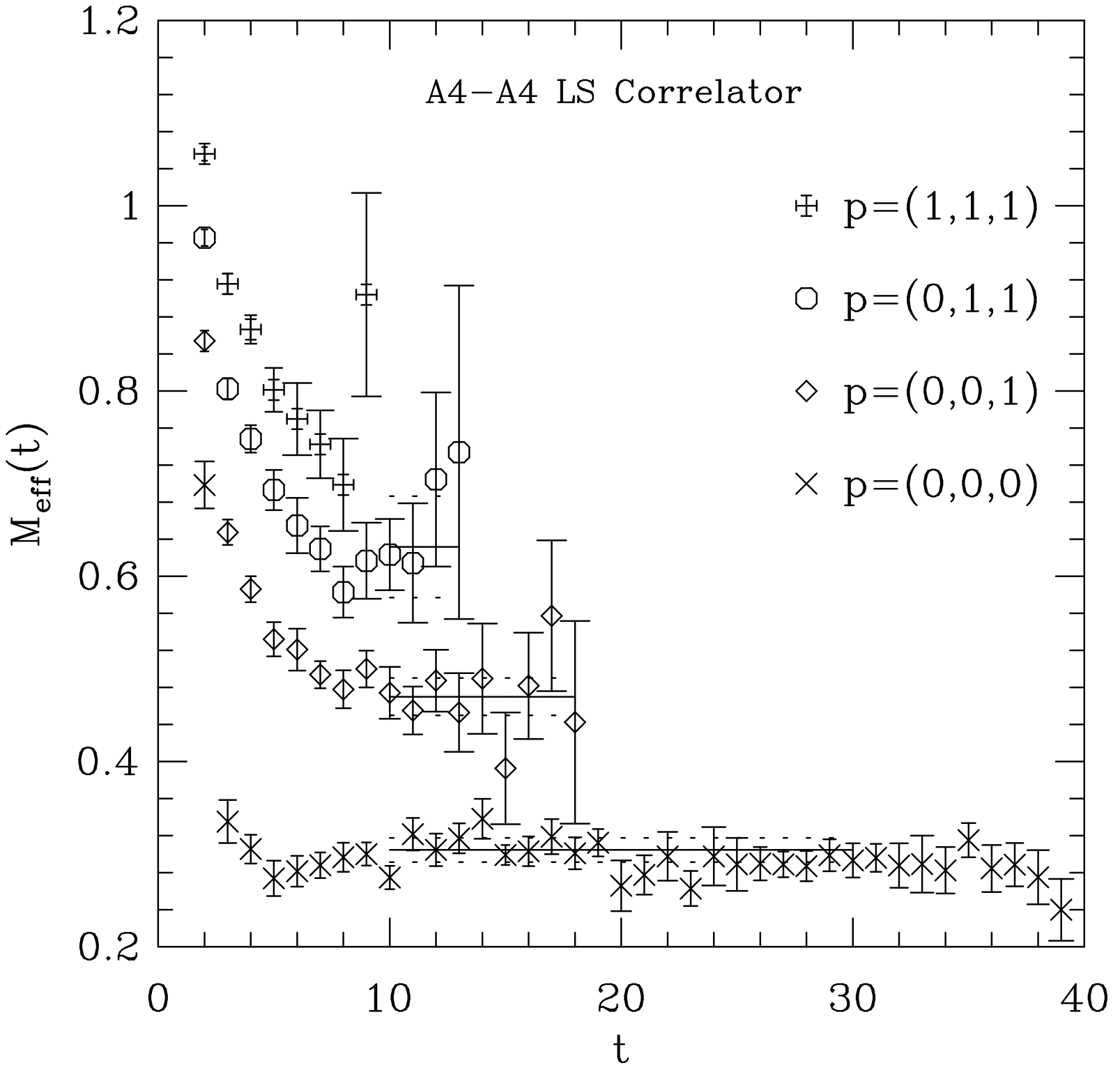

**Fig. 2a.** The $m_{\rm eff}(t)$ plot for $\pi_2\pi_2$ $LS$ correlator for the 4 values of momentum, $\vec{p}=(0,0,0)$, $(0,0,1)$, $(0,1,1)$, and $(1,1,1)$, at $\kappa=0.155$.

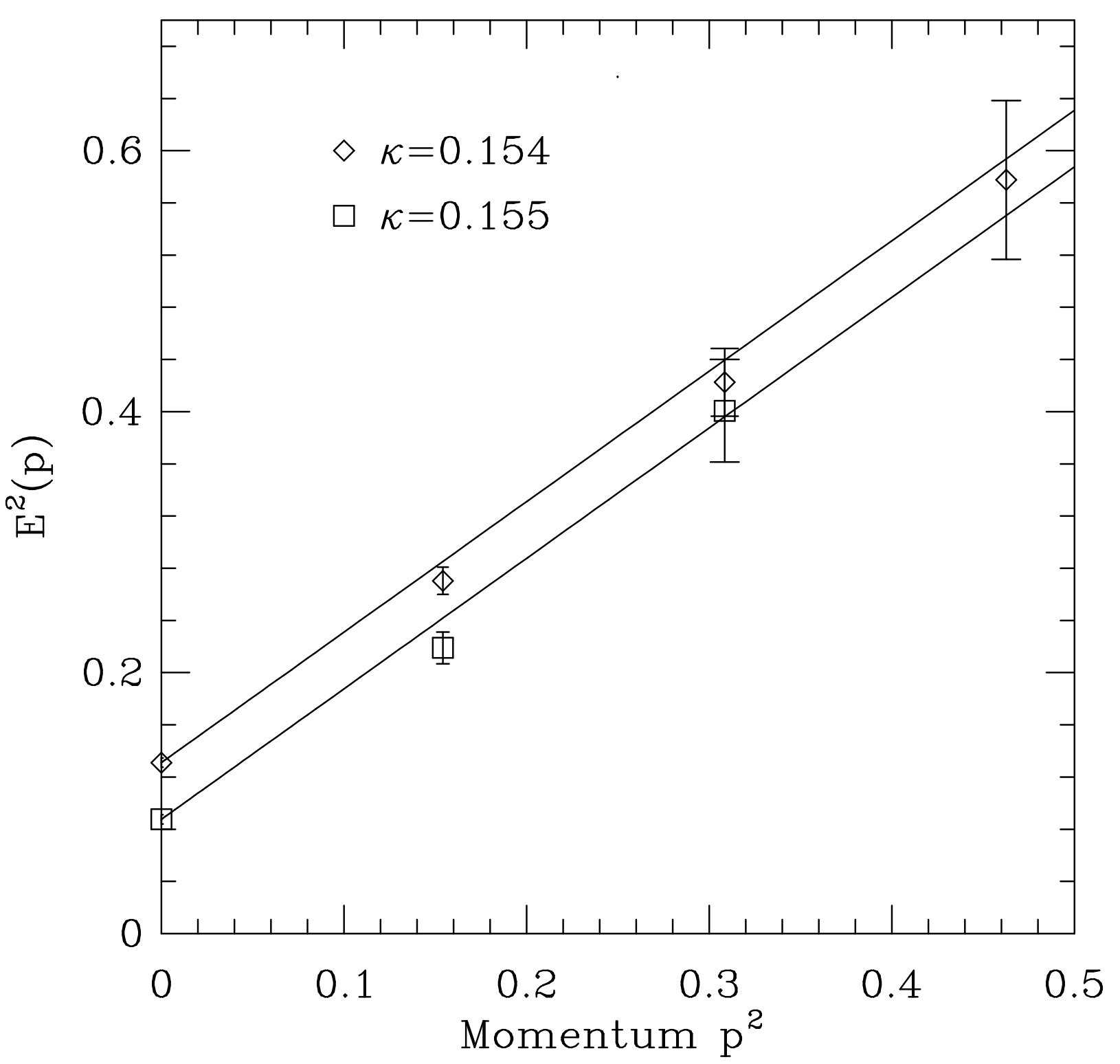

**Fig. 2b.** Comparison of lattice pion spectrum with the continuum dispersion relation $E^2=m^2+p^2$. Calculations were done only for the four lowest values of momentum. We do not quote a result at $\kappa=0.155$ for $\vec{p}=(1,1,1)$ as there is no credible plateau in $m_{\rm eff}(t)$ (see Fig. 2a).

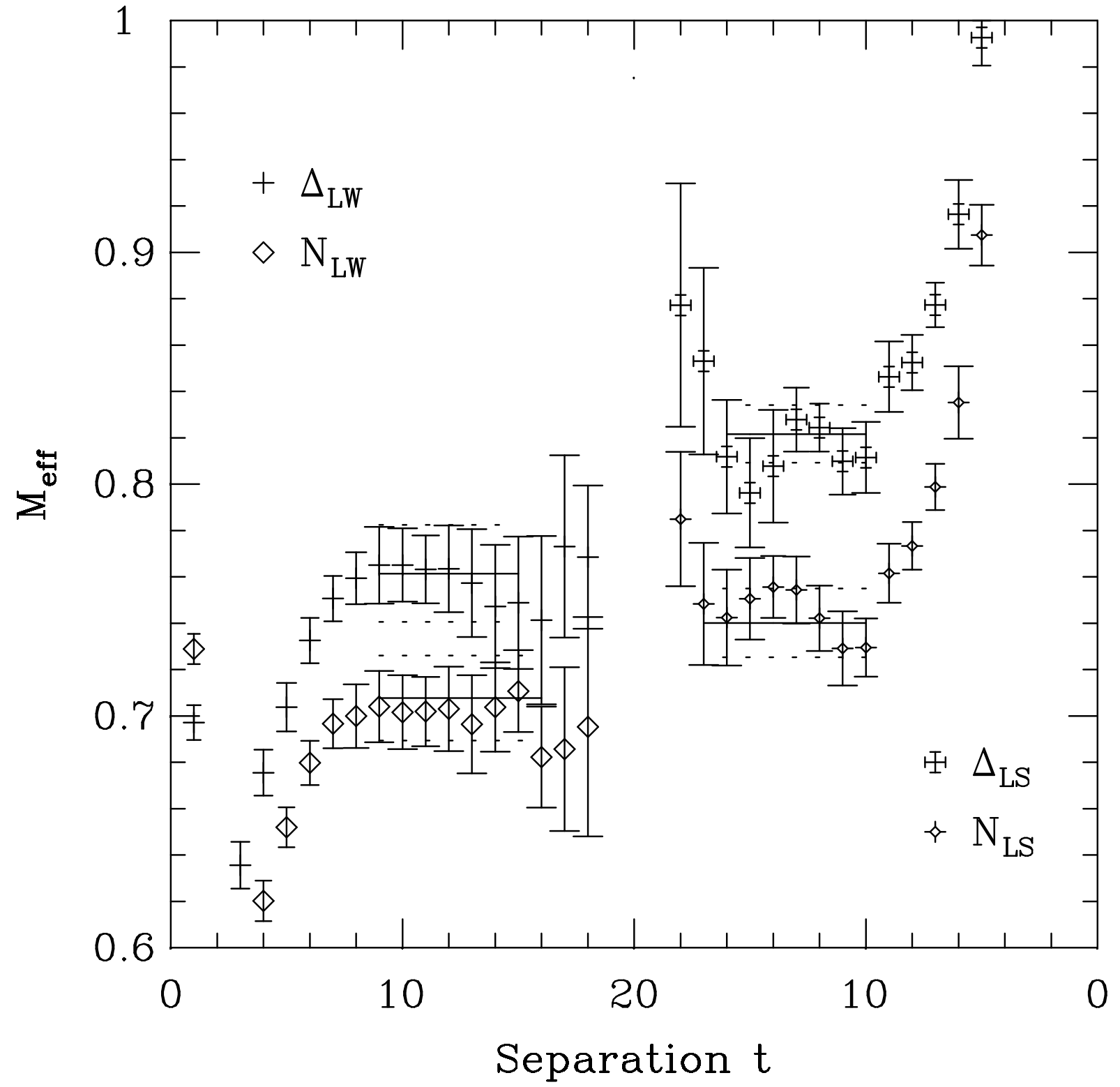

**Fig. 3.** Comparison of the $m_{\rm eff}(t)$ plots for the baryons at $\kappa = 0.154$ using $LW$ and $LS$ correlators.

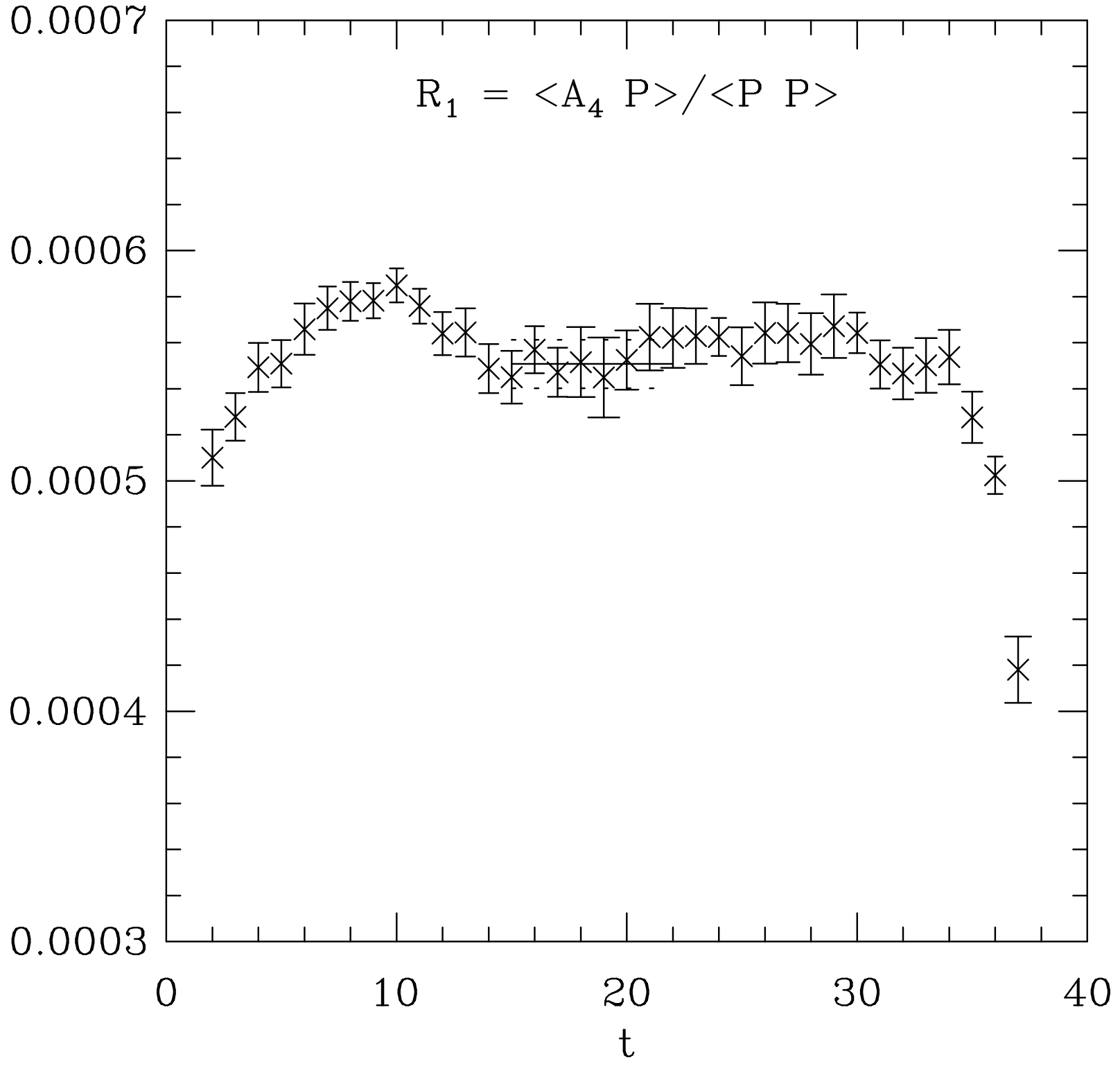

**Fig. 4a.** The ratio of correlators $R_1$ at $\kappa = 0.154$ from which $f_\pi^a$, as defined in the text, is extracted.

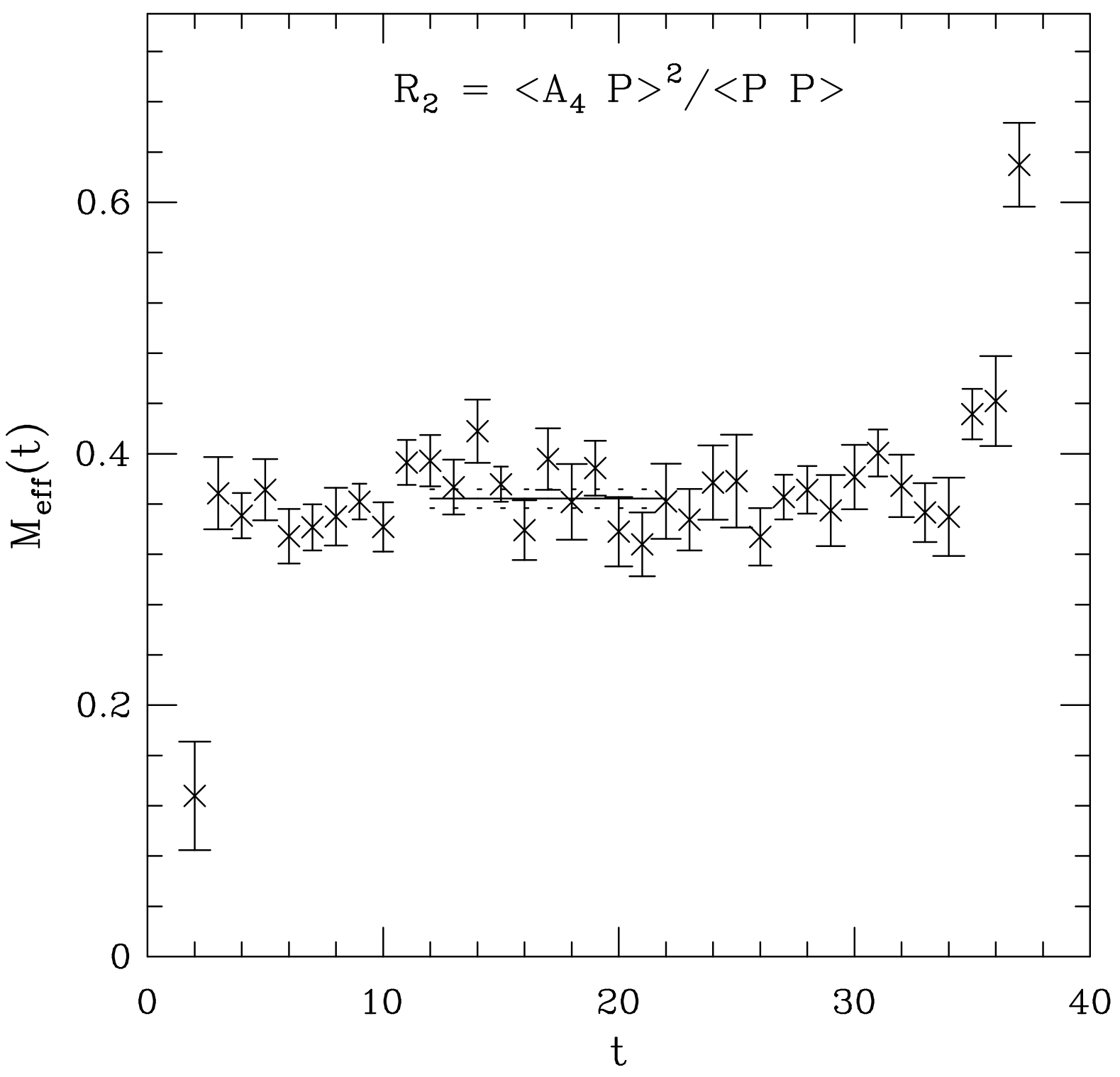

**Fig. 4b.** The $m_{\rm eff}(t)$ plot for the ratio of correlators $R_2$ at $\kappa = 0.154$ from which $f_\pi^c$, as defined in the text, is extracted.

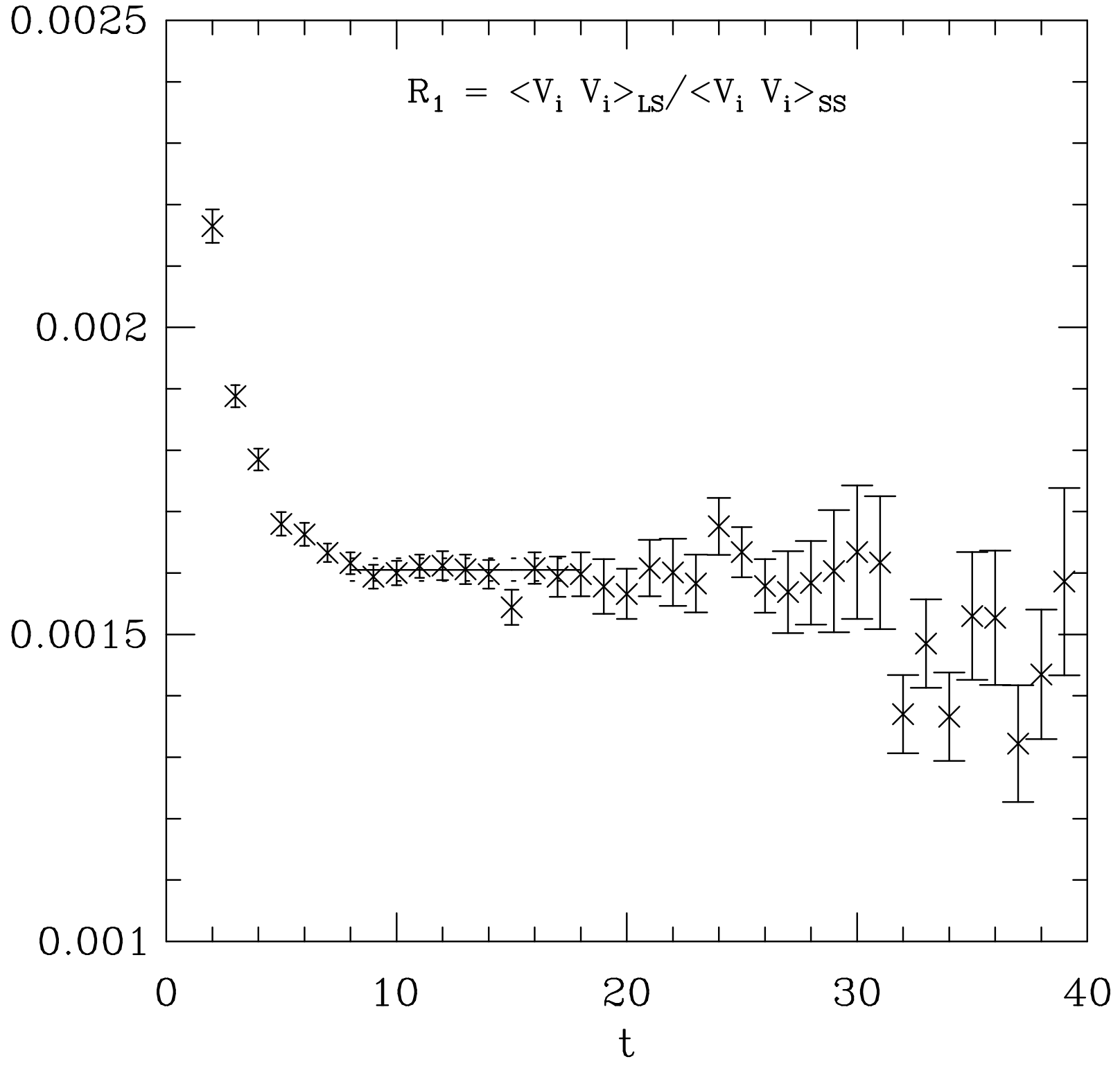

**Fig. 5a.** The ratio of correlators $R_1$ at $\kappa = 0.154$ from which $1/f_V^a$, as defined in the text, is extracted.

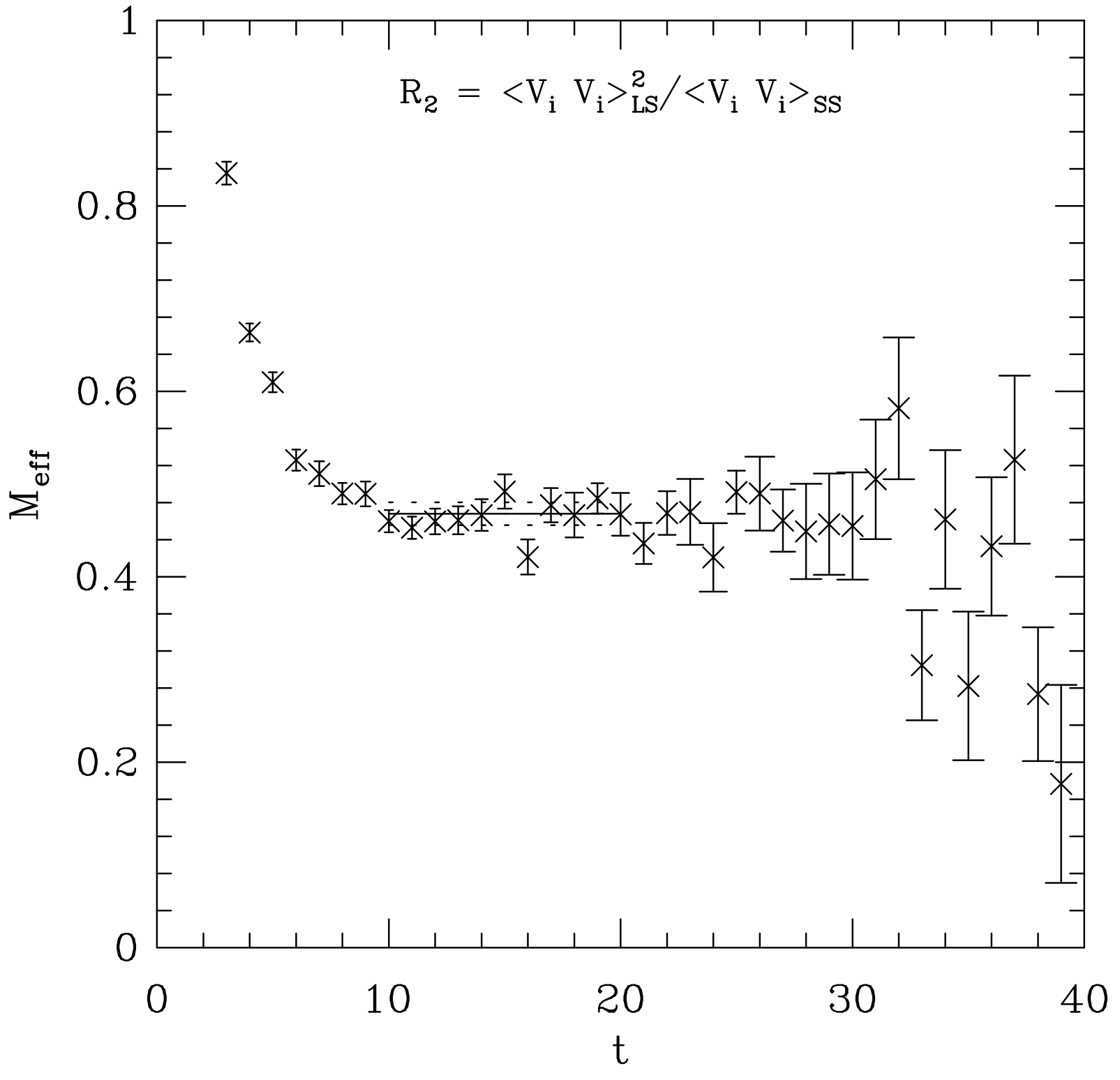

**Fig. 5b.** The $m_{\rm eff}(t)$ plot for the ratio of correlators $R_2$ at $\kappa = 0.154$ from which $1/f_V^b$, as defined in the text, is extracted.

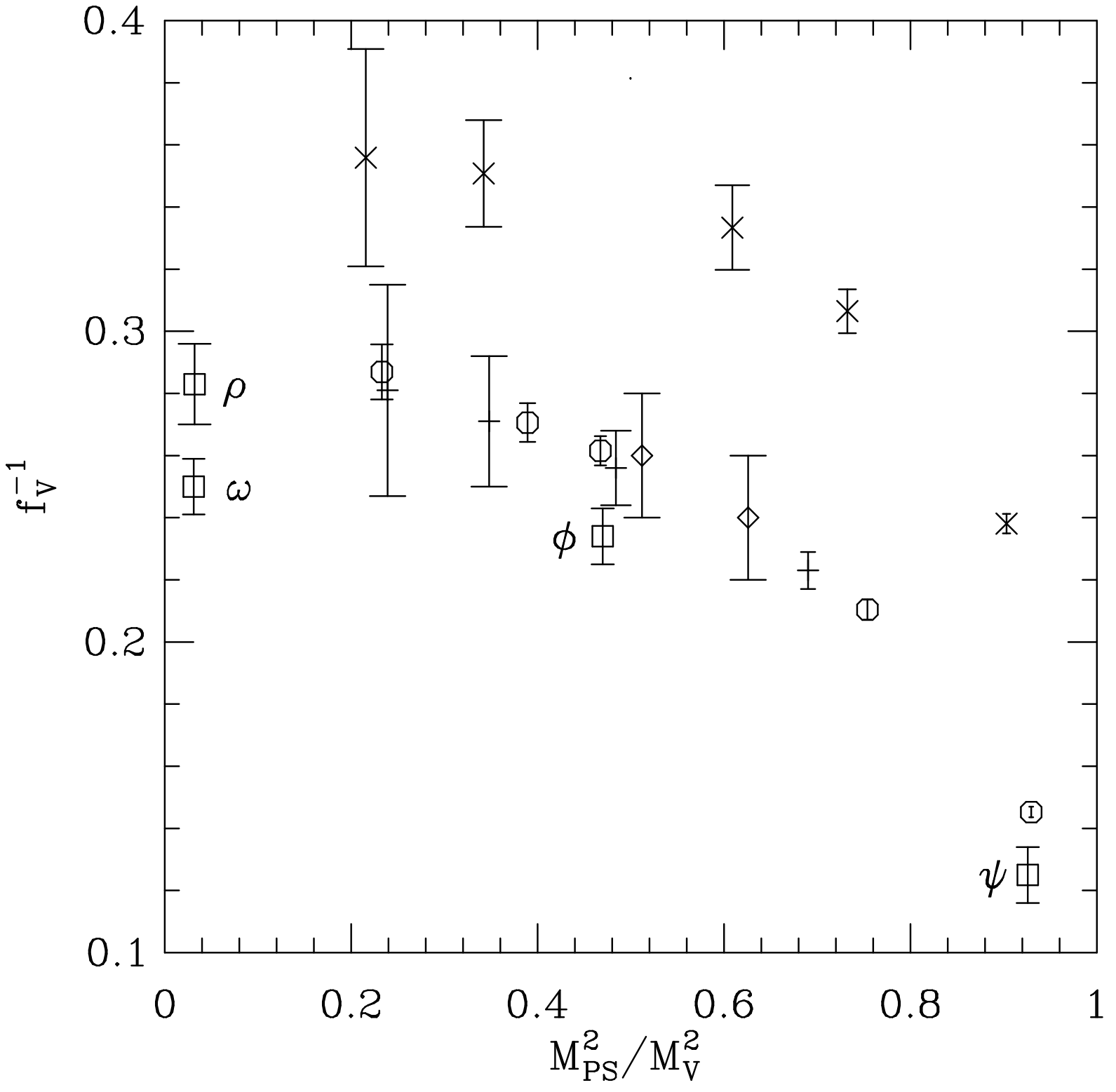

**Fig. 6.** Comparison of the lattice estimates of $1/f_V$ with experimental data. Points labeled by $+$ are the corrected data from Ref. [12], $\circ$ from Ref. [17], $\times$ from Ref. [19] and $\diamond$ from this calculation. Experimental points are indicated by $\square$.

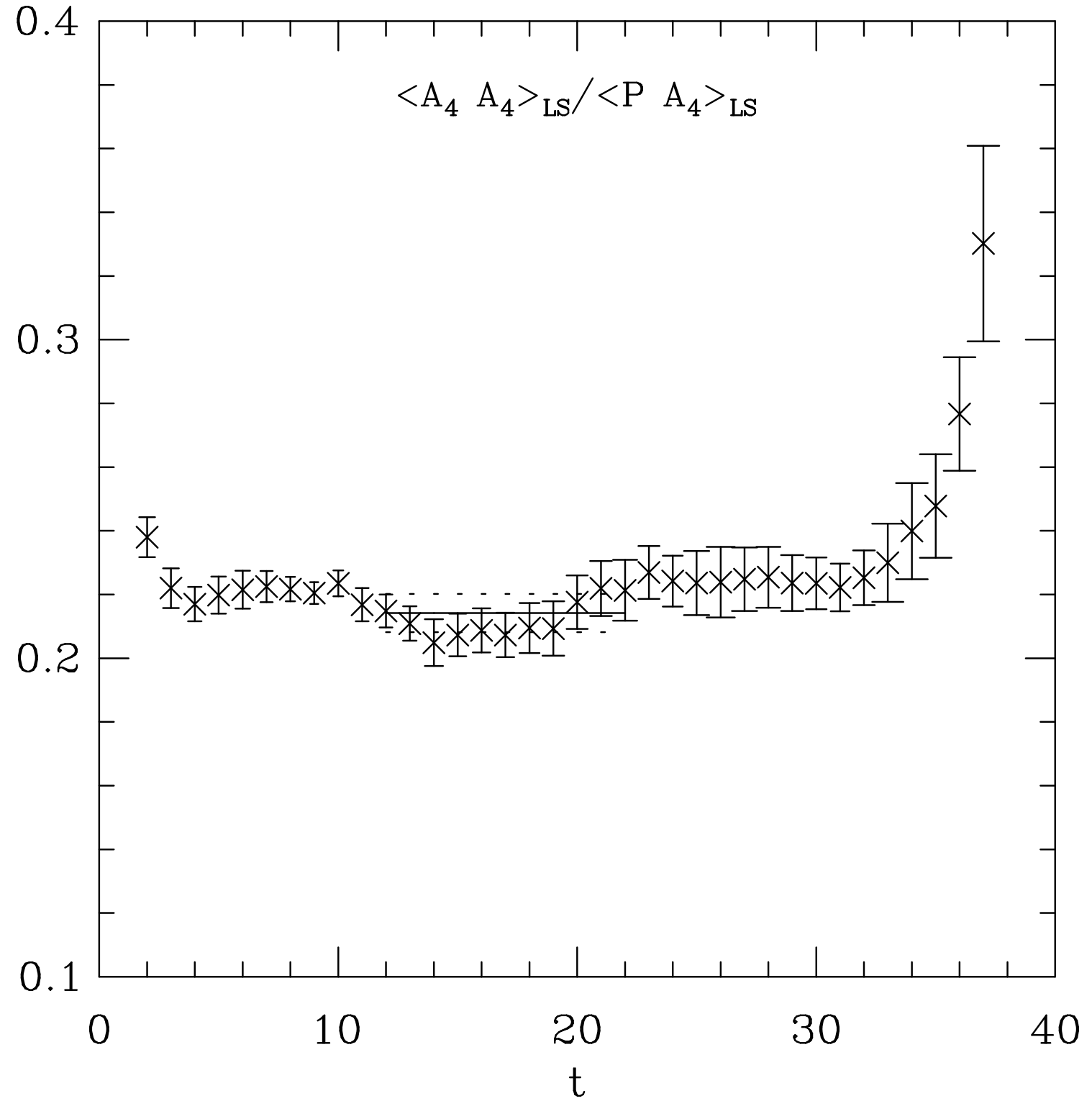

**Fig. 7a.** Plot of the ratio of correlators at $\kappa = 0.155$ from which $m_q^b$, as defined in the text, is extracted.

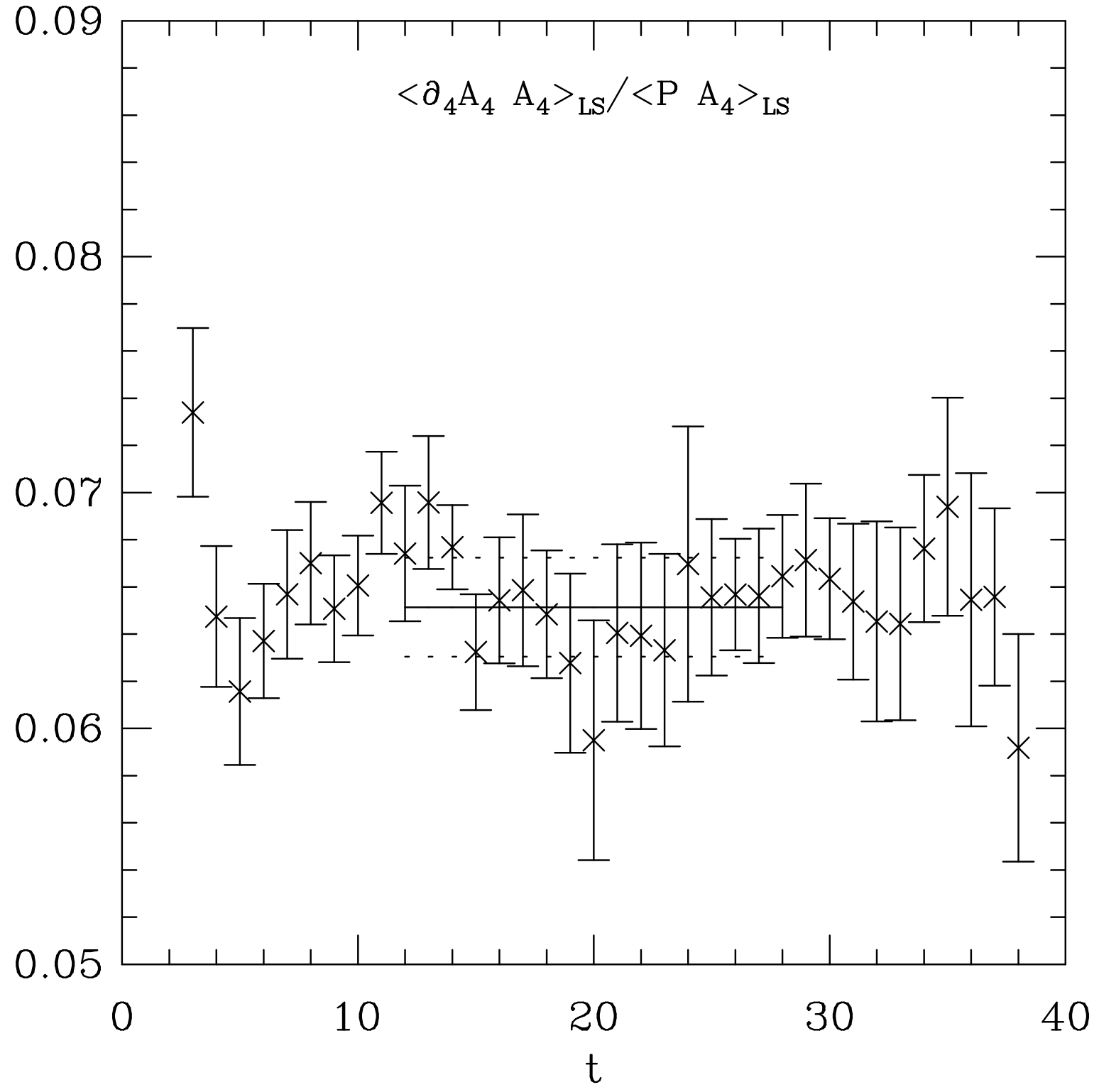

**Fig. 7b.** Plot of the ratio of correlators at $\kappa = 0.155$ from which $m_q^d$, as defined in the text, is extracted.

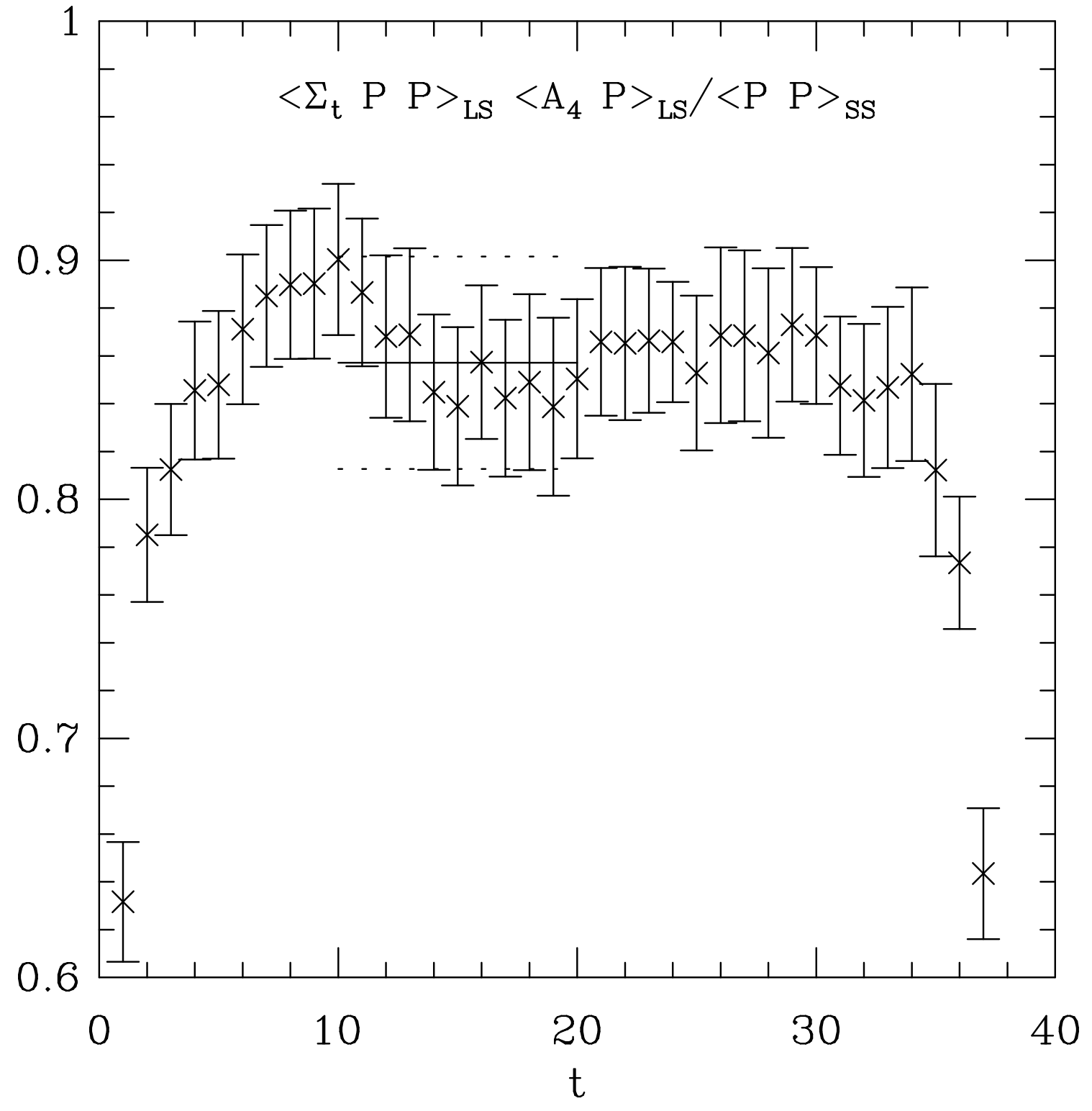

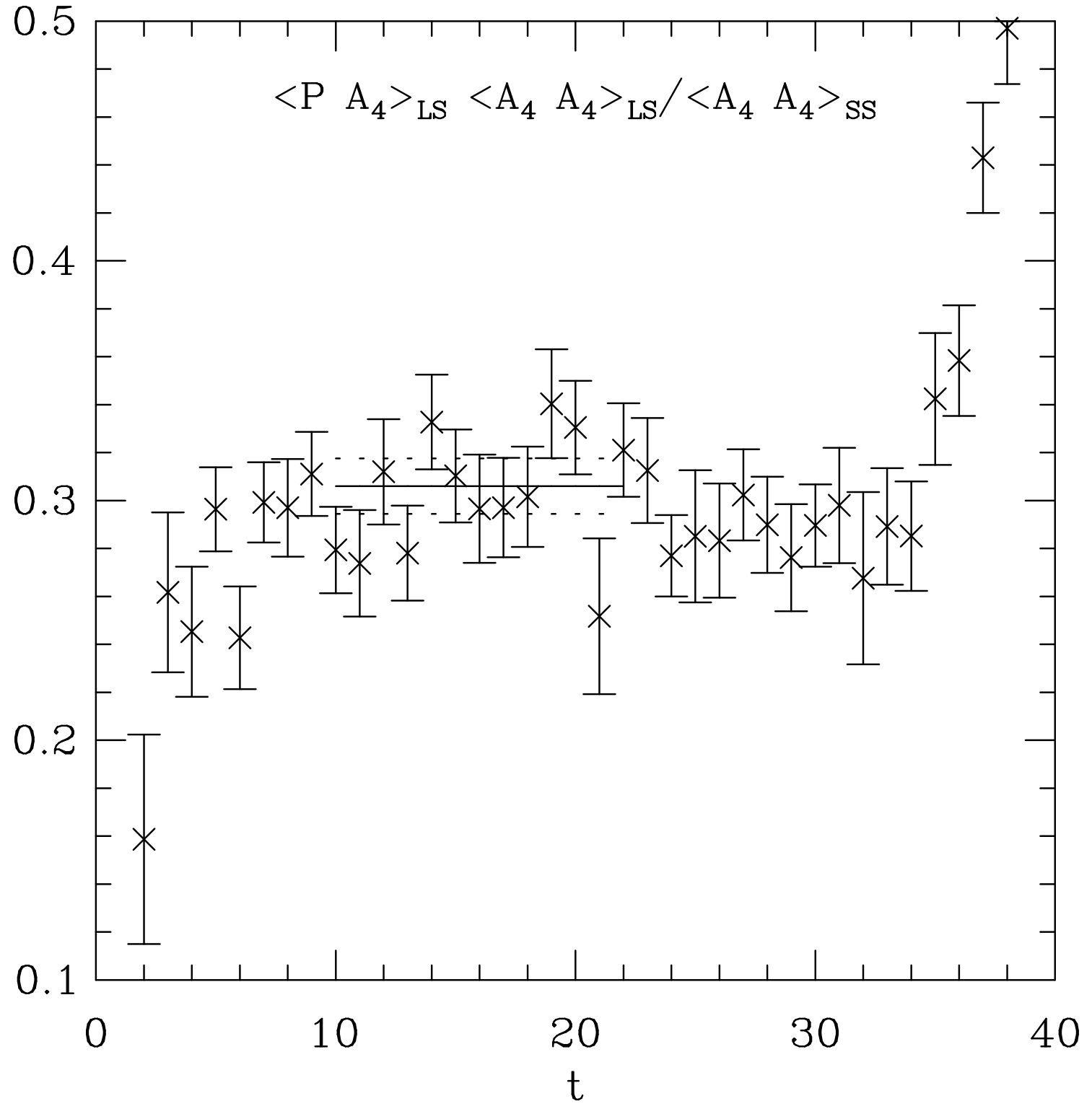

**Fig. 8a.** The plot for the ratio of correlators yielding $\langle\overline{\psi}\psi\rangle^{\widetilde{\mathrm{WI}}}$ at $\kappa = 0.154$. The result has to be multiplied by $4\kappa^2$ to take into account our normalization of quark propagators.

**Fig. 8b.** The $m_{\mathrm{eff}}(t)$ plot for the ratio of correlators yielding $\langle\overline{\psi}\psi\rangle^{\mathrm{GMOR}}$ at $\kappa = 0.155$.

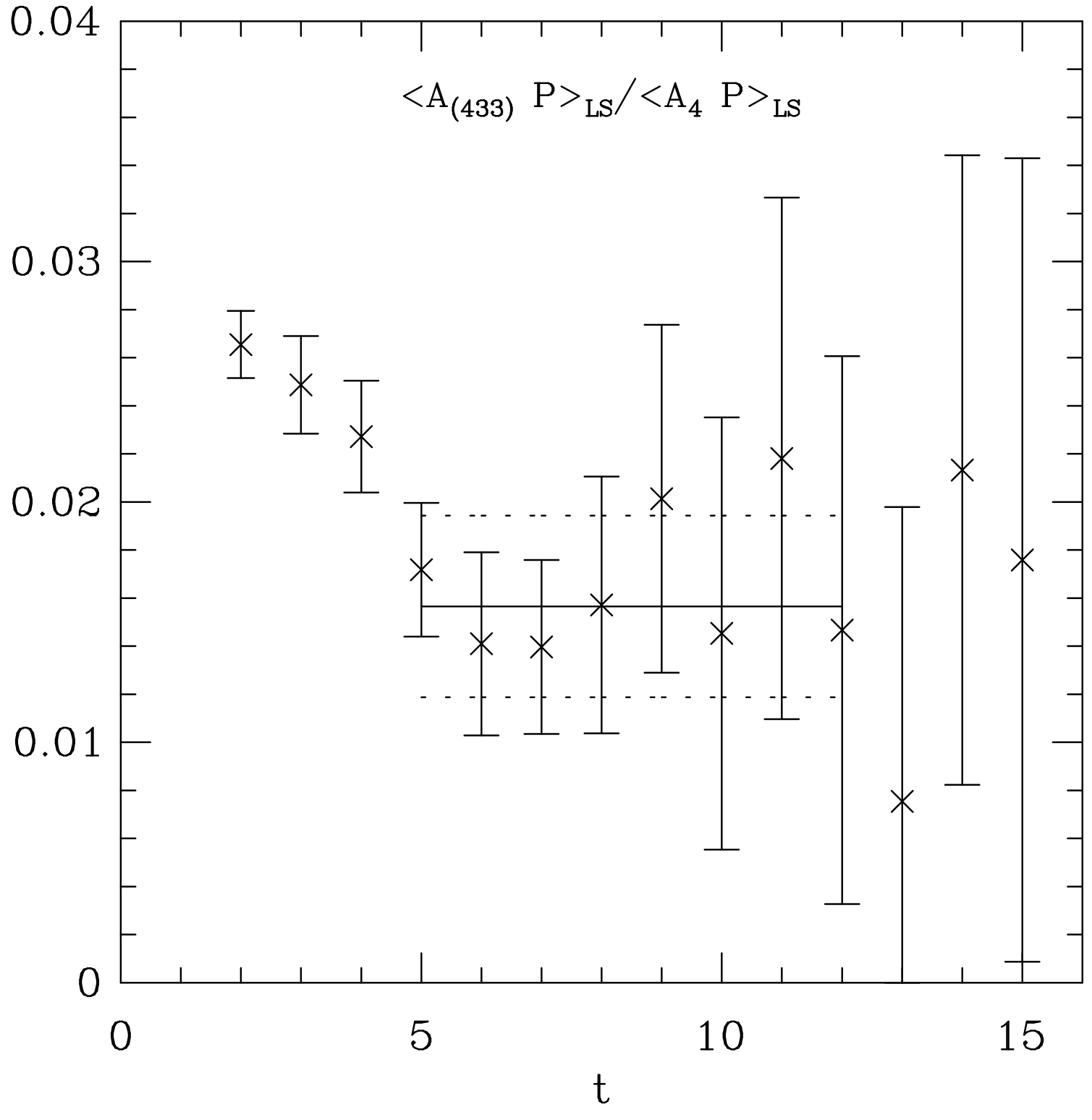

**Fig. 9a.** Plot of the ratio of correlators, $R_{\pi}^{(433)}$, at $\kappa = 0.155$ from which the lattice value of $\langle\xi^2\rangle$ is extracted.

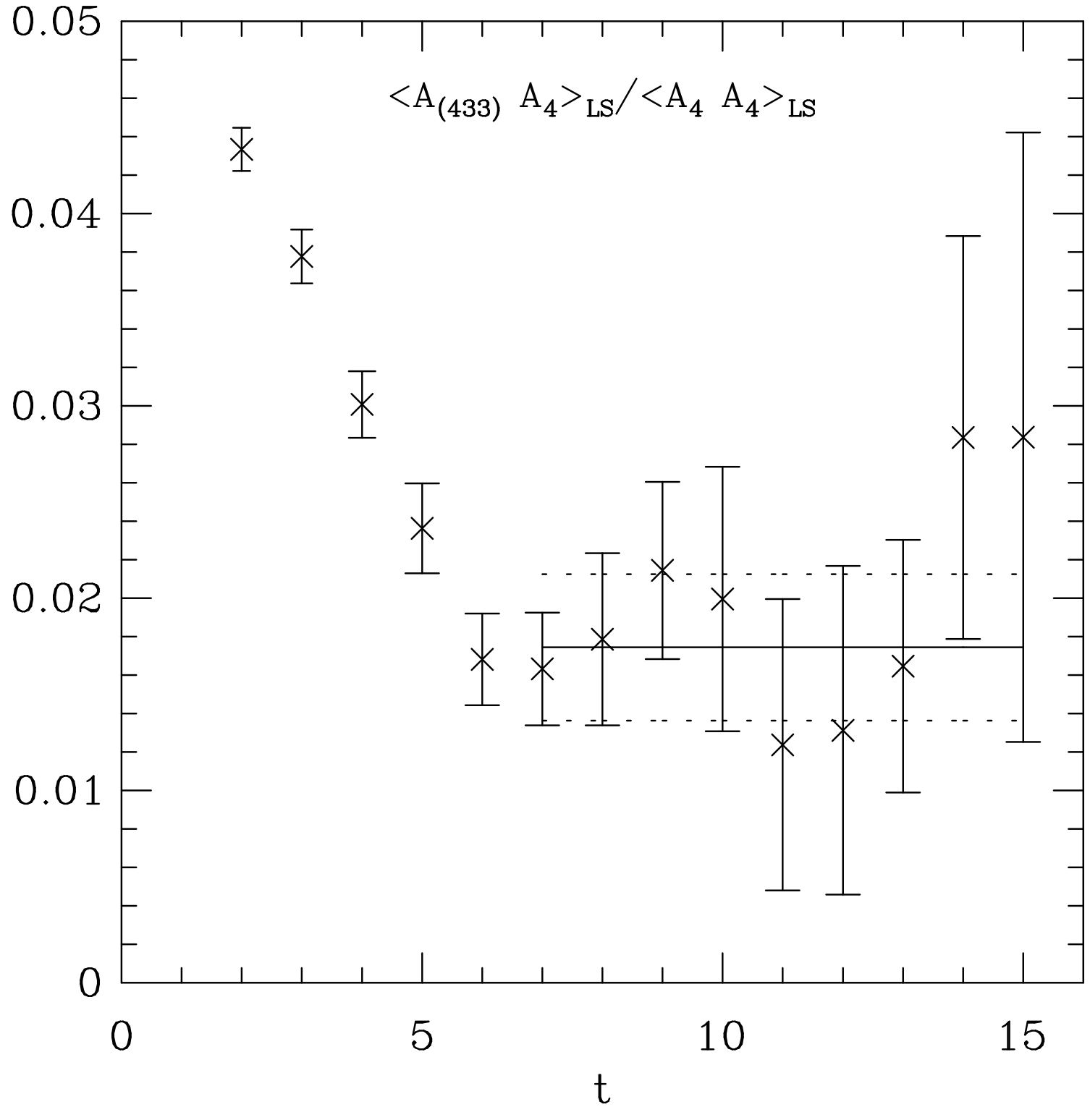

**Fig. 9b.** Plot of the ratio of correlators, $R_{\pi_2}^{(433)}$, at $\kappa = 0.155$ from which the lattice value of $\langle\xi^2\rangle$ is extracted.

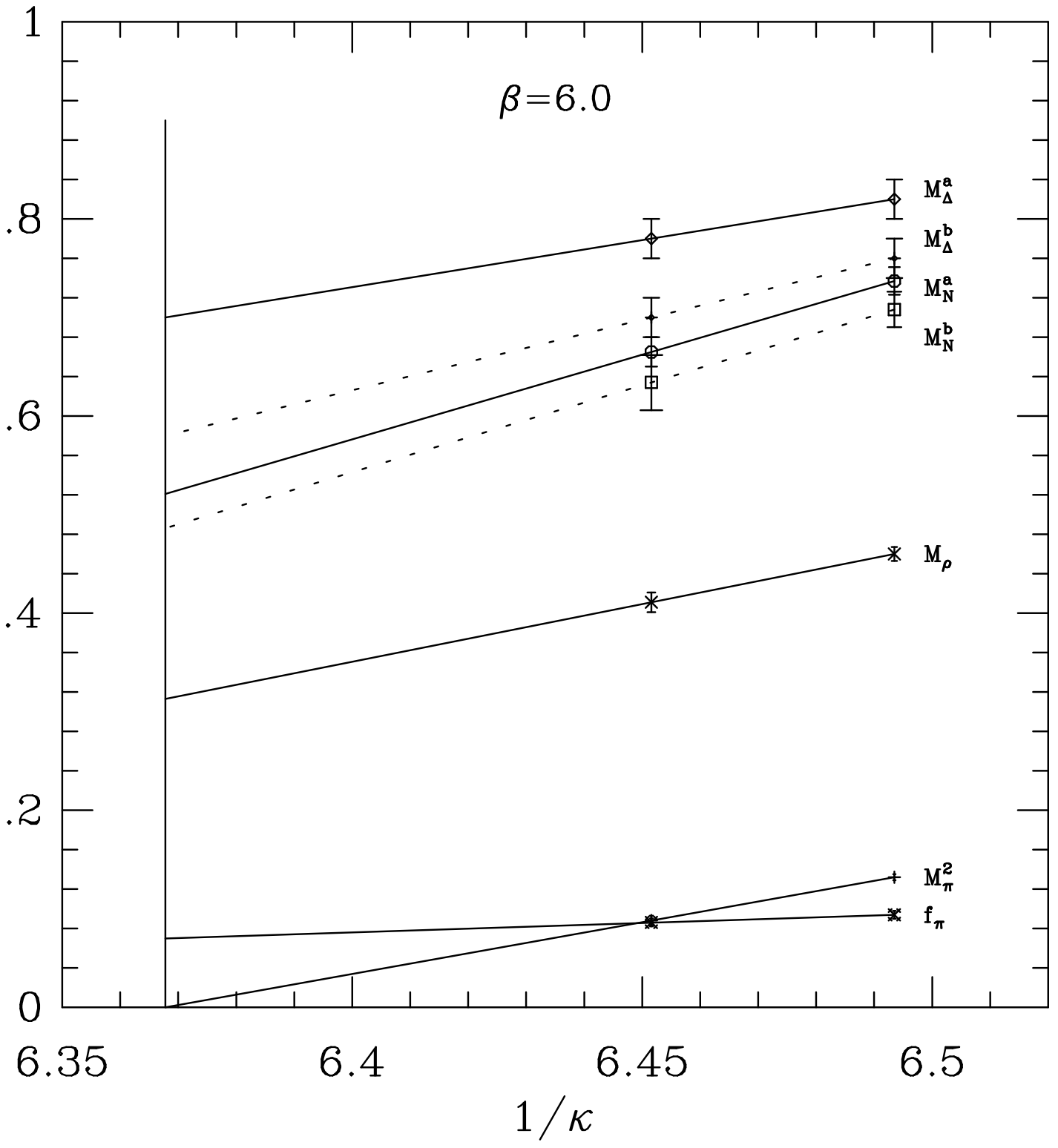

**Fig. 10.** Linear extrapolation of the lattice masses to the chiral limit. We show both sets of baryon masses, those obtained using Wuppertal (superscript $a$) and using wall (superscript $b$) sources.